\documentclass[11pt,a4paper]{article}
\usepackage{jheppub}
\pdfoutput=1
\usepackage{hyperref}
\usepackage[utf8]{inputenc}
\usepackage[T1]{fontenc}
\usepackage{amsmath}
\usepackage{float}
\usepackage{graphicx}
\usepackage{comment}
\usepackage{xcolor}
\DeclareMathAlphabet{\mathpzc}{OT1}{pzc}{mma}{it}
\usepackage{color}
\usepackage{epsfig}
\usepackage{dcolumn}
\usepackage{hyperref}
\usepackage{bm}
\usepackage{amssymb}
\usepackage{slashed}
\usepackage{epstopdf}
\usepackage{ae}
\usepackage[normalem]{ulem}	
\usepackage{nccmath}
\usepackage[nice]{nicefrac}
\pdfoutput=1

\makeatletter
\renewcommand{\maketag@@@}[1]{\hbox{\m@th\normalsize\normalfont#1}}%
\makeatother
\definecolor{red}{rgb}{1,0,0}
\definecolor{darkpink}{rgb}{0.8,0,0.4}

\newcommand{\be}{\begin{equation}}
\newcommand{\ee}{\end{equation}}
\newcommand{\bea}{\begin{eqnarray}}
\newcommand{\eea}{\end{eqnarray}}
\newcommand{\bi}{\begin{itemize}}
\newcommand{\ei}{\end{itemize}}

\newcommand\abs[1]{\left|#1\right|}

\newcommand{\IR}{\textnormal{\tiny \textsc{IR}}}
\newcommand{\UV}{\textnormal{\tiny \textsc{UV}}}

\newcommand{\ew}{\textnormal{\tiny \textsc{EW}}}
\newcommand{\PL}{\textnormal{\tiny \textsc{PL}}}

\newcommand{\B}{\textnormal{\tiny \textsc{B}}}
\newcommand{\cf}{f}

\newcommand{\cF}{F}

\DeclareRobustCommand{\Eq}[1]{Eq.~(\ref{#1})}
\DeclareRobustCommand{\Fig}[1]{Fig.~\ref{#1}}
\DeclareRobustCommand{\Ref}[1]{Ref.~\cite{#1}}
\DeclareRobustCommand{\Sec}[1]{Sec.~\ref{#1}}

\DeclareRobustCommand{\Tab}[1]{Table~\ref{#1}}

\title{A warped relaxion}

\author[a]{Nayara~Fonseca,}
\author[a]{Benedict~von~Harling,}
\author[b]{Leonardo~de~Lima}
\author[c]{and Camila~S.~Machado}
\affiliation[a]{DESY, Notkestrasse 85, 22607 Hamburg, Germany}
\affiliation[b]{Universidade Federal da Fronteira Sul, Av. Edmundo Gaievski 1000, 85770-000 Realeza, Brazil}
\affiliation[c]{PRISMA Cluster of Excellence and Mainz Institute for Theoretical Physics, Johannes Gutenberg-Universit{\"a}t Mainz, 55099 Mainz, Germany}

\emailAdd{nayara.fonseca@desy.de}
\emailAdd{benedict.von.harling@desy.de}
\emailAdd{leonardo.de.lima@uffs.edu.br}
\emailAdd{camachad@uni-mainz.de}

\abstract{We construct a UV completion of the relaxion in a warped extra dimension. We identify the relaxion with the zero mode of the fifth component of a bulk gauge field and show how hierarchically different decay constants for this field can be achieved by different localizations of anomalous terms in the warped space. This framework may also find applications for other axion-like fields. The cutoff of the relaxion model is identified as  the scale of the IR brane where the Higgs lives, which can be as high as $10^6\,$GeV, while above this scale warping takes over in protecting the Higgs mass.}
\preprint{DESY 17-228 ~~ MITP 17-103 }

\begin{document}

\maketitle


\section{Introduction}

The traditional paradigms to approach the hierarchy problem of the Standard Model require new physics close to the electroweak scale, attributing  the smallness of the  Higgs mass  to a symmetry protection (e.g.~supersymmetry) or to the lowering of the cutoff of the theory (e.g.~technicolor).
This class of solutions has been a guide to model building of physics beyond the Standard Model for many years and one of the leading motivations of searches for new physics at the LHC.
An alternative possibility does not predict new physics at the TeV scale, but instead requires multiple vacua with a large range of possible values of the Higgs mass and a selection mechanism such that we end up in the vacuum where the Higgs is light. Recently, a  new dynamical selection mechanism was proposed, the cosmological relaxation of the electroweak scale \cite{Graham:2015cka} (see also \cite{Espinosa:2015eda,  Patil:2015oxa, Jaeckel:2015txa, Gupta:2015uea, Batell:2015fma, Matsedonskyi:2015xta, Marzola:2015dia,DiChiara:2015euo, Evans:2016htp, Kobayashi:2016bue, Evans:2017bjs}).
It relies on the scanning of the Higgs mass parameter by a new field, the relaxion, and a back-reaction mechanism that is triggered when the vacuum expectation value (VEV) of the Higgs has reached the electroweak scale, making the relaxion evolution stop.\footnote{See also $N$naturalness \cite{Arkani-Hamed:2016rle}, where instead of multiple vacua, many copies of the Standard Model are considered to explain the smallness of the electroweak scale. The way reheating behaves is such that only the copy with the smallest Higgs mass is efficiently reheated.
} This is a radical change of paradigm as it implies that the naturalness problem of the  Standard Model ceases to be a reason to expect new physics close to the TeV scale.

In what follows we review  the relaxation mechanism for which an axion-like scalar $\phi$ is introduced which couples to the Higgs doublet $H$ via the potential
\be \label{eq:VRelaxion1a}
 V(\phi, H) \, \supset \, -\,\left(\Lambda^2 \,-\, g'  \Lambda \, \phi\right) H^2 \,+ \lambda\, H^4 + \, g  \Lambda^3  \phi \, + \, \Lambda_f^4(H) \cos{\left(\frac{\phi}{f}\right)} \, .
\ee
Here $\Lambda$ is the cutoff which sets the Higgs mass parameter, $f$ the decay constant of the relaxion, $\lambda$ the Higgs quartic coupling, $g$ and $g'$ are small dimensionless couplings, and $\Lambda_f(H)$ is a scale which depends on the Higgs VEV.
Assuming a classical time evolution with slow-roll conditions, the second-last term in \Eq{eq:VRelaxion1a} causes the relaxion to move downwards following its potential.
The effective Higgs mass parameter in the $\phi$ background, the first term in parenthesis in \Eq{eq:VRelaxion1a}, then varies accordingly.
The relaxion is assumed to start with a VEV such that this mass parameter is initially positive.
Due to the evolution of the relaxion, the mass parameter then eventually turns tachyonic, triggering electroweak symmetry breaking. In the presence of a Higgs VEV, the oscillatory barrier from the last term grows, until its slope matches the slope of the linear term. For technically natural parameters in the potential, this causes the relaxion to stop once the Higgs VEV has reached the electroweak scale.
There must be some mechanism to dissipate the kinetic energy of the relaxion during its evolution such that the field does not overshoot the barriers. If the dynamics happens during a period of inflation, Hubble friction can provide the dissipation necessary to slow down the field \cite{Graham:2015cka}. As an alternative to inflation,  one can also consider friction due to particle production as proposed in \Ref{Hook:2016mqo} or  finite temperature effects in the early universe as in \Ref{Hardy:2015laa}.

Note that  the linear terms in $\phi$ are in conflict with the assumption that the relaxion is  a pseudo-Nambu-Goldstone boson as they  explicitly break the axion shift symmetry \cite{Gupta:2015uea}. This may be reconciled if the linear terms arise from a second oscillatory potential with a period much larger than $f$. This is realized if the  potential takes the form  \cite{Choi:2015fiu, Kaplan:2015fuy, Fonseca:2016eoo}:\footnote{See also Refs.~\cite{Kim:2004rp, Harigaya:2014eta, Choi:2014rja, Higaki:2014pja, Harigaya:2014rga, Peloso:2015dsa} for similar earlier ideas in inflation  model building. For the viability of the relaxation mechanism in string theory in the context of axion monodromy, see Ref.~\cite{McAllister:2016vzi}.}
\be \label{eq:VRelaxion2a}
V(\phi,H) \, \supset \, - \, \Lambda^2 H^2 \,+ \lambda\, H^4 \, + \, \Lambda_F^4(H) \cos{\left(\frac{\phi}{F}\right)} \,+ \, \Lambda_f^4(H) \cos{\left(\frac{\phi}{f}\right)} \, ,
\ee
where $F \gg f$ is another decay constant and $\Lambda_F(H)$ another scale that depends on the Higgs in such a way as to reproduce the second and fourth term in \Eq{eq:VRelaxion1a} after expanding in $\phi/F$. An interesting possibility to obtain this type of potential is the clockwork construction which was first realized for axion-like fields in Refs.~\cite{Choi:2015fiu,Kaplan:2015fuy} and generalized for applications other than the relaxion in \Ref{Giudice:2016yja}. Further developments regarding the 5D continuum limit of the clockwork can be found in Refs.~\cite{Ahmed:2016viu, Craig:2017cda, Giudice:2017suc, Choi:2017ncj}. Besides the clockwork, one can also generate a potential of the form in \Eq{eq:VRelaxion2a} in realizations inspired by dimensional deconstruction \cite{Hill:2000mu, ArkaniHamed:2001ca}, as in \Ref{Fonseca:2016eoo}.

\begin{figure}
\center
\includegraphics[scale=1.0]{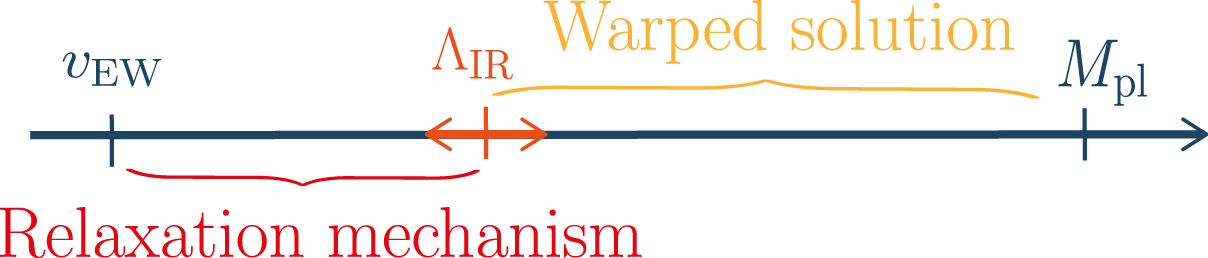}
\caption{\label{fig:hierarchy} \emph{A UV completion for the relaxion model is obtained by embedding the relaxion and the Higgs into a warped extra dimension. The hierarchy problem is then solved in two steps: the relaxation mechanism protects the Higgs mass up to the IR scale (which can be much larger than the electroweak scale) and from there warping provides protection till the Planck scale. }}
\end{figure}

In this work, we show how the required potential for the relaxation mechanism to work can be naturally obtained by embedding the relaxion and Higgs into a warped extra dimension. We consider a slice of AdS$_5$ space which is bounded by two branes, as in the Randall-Sundrum model \cite{Randall:1999ee}. However, in our setup the IR scale or warped-down AdS scale is not of order TeV but can be much larger.
We introduce a $U(1)$ gauge field in the bulk of the extra dimension and break the gauge symmetry on the two branes. The 5th component $A_5$ of the gauge field then gives rise to one massless scalar mode in 4D which we identify with the relaxion. In order to generate a potential, we introduce anomalous couplings of $A_5$ to two non-abelian gauge groups.
The wavefunction of the massless mode from $A_5$ is exponentially peaked towards the IR brane (see e.g.~\cite{Contino:2003ve, Hosotani:2005nz, Falkowski:2006vi}). Depending on where the anomalous terms are localized, this can yield a large hierarchy between the decay constants for the couplings of the relaxion to the gauge groups. We assume that the gauge groups confine at energies below the compactification scale. Instantons then generate periodic potentials for the relaxion as in \Eq{eq:VRelaxion2a} with periods given by the decay constants.\footnote{A potential for $A_5$
can be generated perturbatively if the underlying gauge field is coupled to charged bulk states. In the non-abelian case (see e.g.~\cite{Contino:2003ve}), this includes the gauge fields themselves due to the non-linear interactions, while the abelian case requires charged scalars or fermions in the bulk (see e.g.~\cite{ArkaniHamed:2003wu}). Here we consider a $U(1)$ gauge field and do not add charged bulk states as we are interested in generating a non-perturbative potential for $A_5$.} Due to the warping, these periods can thus naturally be hierarchically different as required. We embed the Higgs at or near the IR brane. Its mass parameter is then naturally of order the IR scale which we identify with the cutoff of the relaxion theory. The required Higgs-relaxion couplings can be obtained by introducing fermions on the IR brane with higher-dimensional or Yukawa couplings to the Higgs. To summarize, the warping does two things: Firstly, it generates the hierarchy between the decay constants $F$ and $f$ in \Eq{eq:VRelaxion2a} and thereby explains the smallness of the couplings $g$ and $g'$ in \Eq{eq:VRelaxion1a}. Secondly, it provides a UV completion\footnote{As a caveat, we should stress that the Randall-Sundrum model itself requires a UV completion. In particular, near the IR brane gravity becomes strongly coupled at energies not far above the IR scale. Near that brane, the UV completion therefore needs to kick in at correspondingly low scales. There are known UV completions to the Randall-Sundrum model in string theory \cite{Klebanov:2000hb,Giddings:2001yu}. }  for the relaxion. The relaxation mechanism protects the Higgs up to the IR scale above which warping takes over.\footnote{See \cite{Batell:2015fma,Evans:2016htp,Evans:2017bjs} for how the relaxation mechanism can protect the Higgs up to some high supersymmetry-breaking scale instead.
}
We illustrate this in \Fig{fig:hierarchy}. Alternatively, one can think of the relaxation mechanism in our construction as a solution to the little hierachy problem of Randall-Sundrum models.\footnote{See \cite{Gherghetta:2011wc} for an alternative solution where an accidental form of supersymmetry protects a little hierarchy between the electroweak scale and the IR scale of the Randall-Sundrum model.} As is well-known, various experimental constraints (the most stringent ones coming from CP violation in $K-\bar{K}$-mixing and the electirc dipole moment of the neutron) require that the IR scale in these models is of order 10 TeV or above. This means that a residual tuning in the permille range is necessary to generate the electroweak scale. In our construction with warping and the relaxion, on the other hand, no such tuning is required.

We find that for an effective anomalous coupling localized on the UV brane, the decay constant is
of order $M_\PL^2/\Lambda_\IR$ with $M_\PL$ and $\Lambda_\IR$ being the Planck and IR scale. For an anomalous coupling in the bulk, we instead find a decay constant of order $\Lambda_\IR$. We then identify $F=M_\PL^2/\Lambda_\IR$ and $f=\Lambda_\IR$.
Generating a suitable barrier $\Lambda_f^4(H) \cos(\phi/f)$ for the relaxion requires some additional structure. The reason is that this term generically contains a contribution which is independent of the Higgs and which could stop the relaxion before the Higgs VEV has reached the electroweak scale. To avoid this problem, we consider two different options. One employs a construction from \Ref{Graham:2015cka} for which new fermions are introduced which couple to the Higgs. If the masses of these fermions are near the electroweak scale, the Higgs-independent barrier can be sufficiently small. The drawback of this construction is a coincidence problem as it requires to introduce the fermions at a scale which is dynamically generated by the relaxation mechanism and thus a priori determined by completely different parameters. An interesting alternative is the so-called double-scanner mechanism of \Ref{Espinosa:2015eda} (see also \cite{Evans:2016htp}).
To this end, one introduces another axion-like scalar which dynamically cancels off the Higgs-independent barrier. We identify this axion-like scalar with the 5th component of another $U(1)$ gauge field in the bulk of the extra dimension. We then  show how the potential which is required for the double-scanner mechanism can be obtained. This construction is largely independent of the embedding into warped space and can therefore also be useful for other UV completions of the relaxion. For both options to generate the barrier, we discuss the relevant theoretical and phenomenological constraints for successful relaxation. The highest cutoff and IR scale consistent with these constraints in our warped implementation of the relaxation mechanism is $\Lambda=\Lambda_\IR\lesssim  10^6 \,$GeV.

The plan of this work is as follows. In \Sec{sec:HierarchicalDecayConstants}, we discuss the properties of the $A_5$ and show how hierarchical decay constants can be obtained. In \Sec{sec:RelaxionPotentialFromA5}, we generate the desired potential for the relaxation mechanism. We analyse the relevant constraints to guarantee a successful relaxation of the electroweak scale  in \Sec{sec:constraints}. In \Sec{sec:doublescanner}, we present our implementation of the double-scanner mechanism and we conclude in \Sec{sec:conclusion}.
Additional details are given in three appendices.

\section{Hierarchical decay constants from warped space}
\label{sec:HierarchicalDecayConstants}
We will now show how hierarchical decay constants can be obtained from warped space. These will be used in later sections to generate the relaxion potential. We consider a slice of AdS$_5$ space with metric in conformal coordinates given by
\begin{align}
\label{e:metric}
ds^2 \, = \, a^2(z)\,(\eta_{\mu\nu} dx^\mu dx^\nu \, - \, dz^2) \,,
\end{align}
where $a(z)= (kz)^{-1}$ is the warp factor with $k$ being the AdS curvature scale (see e.g.~\cite{Gherghetta:2010cj} for a review). The slice is bounded by the UV brane at $z_\UV= 1/k$ and the IR brane at $z_\IR= e^{kL}/k$. The length $L$ of the extra dimension can be stabilized for example by means of the Goldberger-Wise mechanism \cite{Goldberger:1999uk}.
The effective 4D Planck scale for this space is given by $M_\PL^2 \simeq M_*^3/k$, where $M_*$ is the 5D Planck scale. We will assume that the Planck scale and the AdS scale are of the same order of magnitude (and will later often equate them). For later convenience, let us also define the IR scale
$\Lambda_\IR \equiv k\,e^{-kL}$.

Let us consider a $U(1)$ gauge boson in the bulk. Its action is given by
\be
S_{\rm 5D} \,\supset \, \int   d^4x \, dz \, \sqrt{g} \, \left( -\frac{1}{4 g_5^2} \, F_{MN}F^{MN}  \right) \, ,
\ee
where $F_{MN}$ is the $U(1)$ field strength, $g_5$ the 5D gauge coupling and $\sqrt{g}=a^5(z)$. In order to eliminate the mixing between $A_{\mu}$ and $A_5$, we add the gauge fixing term (see e.g.~\cite{Flacke:2006ad,Contino:2003ve})
\be
S_{\rm 5D} \, \supset \, -\int \! d^4x\,dz \, \sqrt{g} \,\frac{1}{2g_5^2 \xi} \, \left[ g^{\mu \nu} \partial_{\mu}A_{\nu} - \, \frac{g^{55} \xi}{a(z)}\partial_5 (A_5 a(z))\right]^2 \, .
\ee
The bulk equations of motion for the 4D component $A_\mu$ and the 5th component $A_5$ then read
\begin{align}
 \eta^{\mu \sigma} \eta^{\lambda \nu} \left(\partial_\sigma \, F_{\mu \lambda} \, + \, \frac{1}{\xi} \partial_\lambda \partial_\mu A_\sigma \right) \, + \, a(z)^{-1} \partial_5 \Bigl(a(z) \eta^{\mu \nu}\partial_5 A_\mu \Bigr) \, & = \, 0\\
\eta^{\mu \nu} \partial_\mu \partial_\nu \, A_5 \, + \, \xi \partial_5 \Bigl( a(z)^{-1}\partial_5 \bigl( a(z) A_5 \bigr)\Bigr) \, & = \, 0 \, .
\end{align}

\begin{figure}
\center
\includegraphics[scale=0.5]{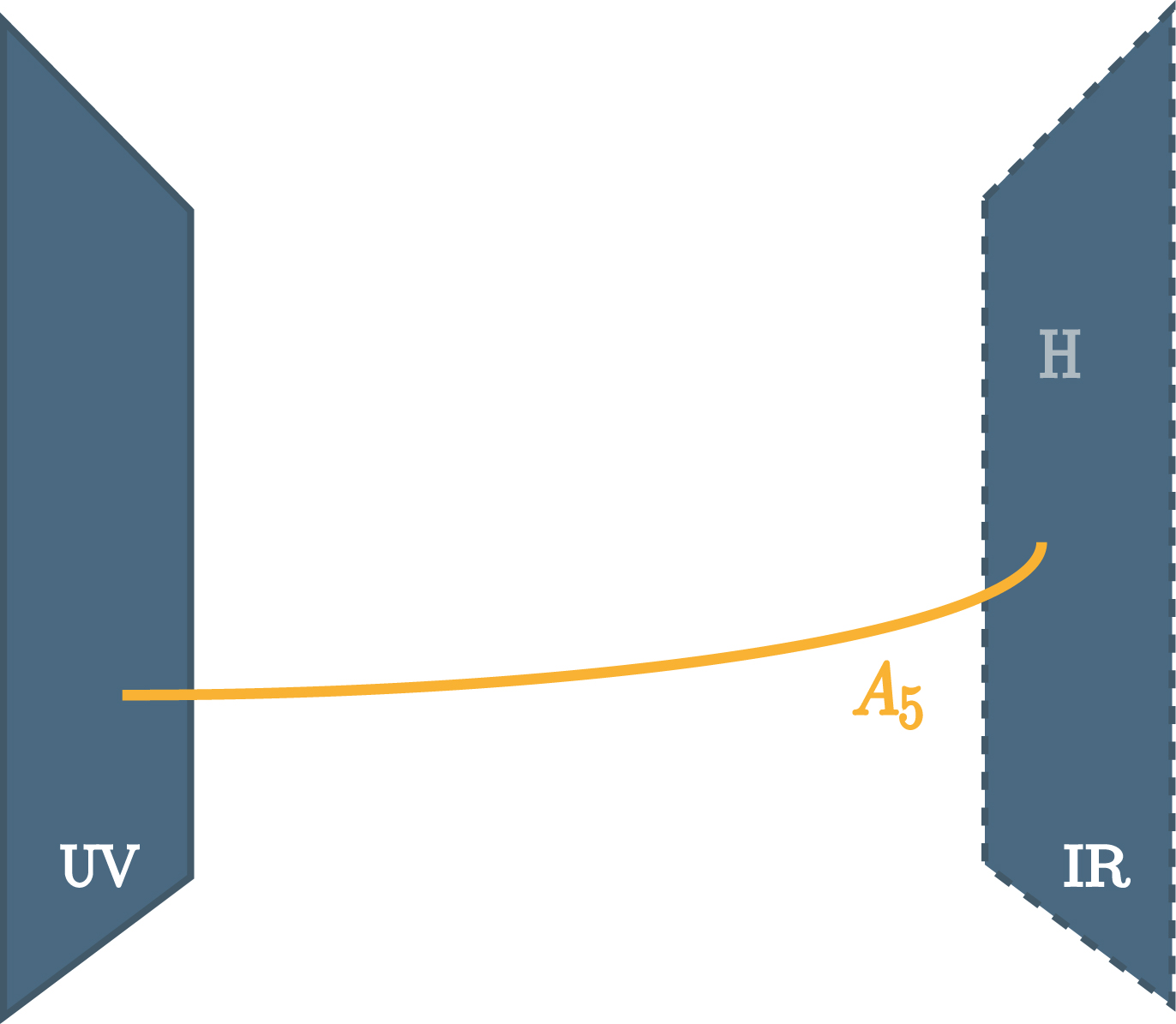}
\caption{\label{fig:warped} \emph{Sketch of a  slice of AdS$_5$ space which is bounded by two branes. We identify the relaxion with the 5th component of a U(1) gauge field in the bulk. Its wavefunction is then localized towards the IR brane. The Higgs is localized on (or near) the IR brane. The UV brane corresponds to the Planck scale. We draw the IR brane with a dashed contour as a reminder that the IR scale in our model can be much larger than the usual TeV scale of the Randall-Sundrum model. }}
\end{figure}

We are interested in obtaining a massless scalar mode from the bulk gauge boson. To this end, we break the gauge symmetry on both branes by imposing Dirichlet boundary conditions on $A_\mu$. For consistency, this then requires to impose Neumann boundary conditions for $A_5$. Together the boundary conditions read
\be
\label{e:bcs}
\left. A_\mu\right\vert_{\UV,\IR} \, =\,  0 \, , \qquad  \left. \partial_5\big(a(z) A_5 \big)\right\vert_{\UV,\IR} \, = \, 0 \,.
\ee
Alternatively we could break the gauge symmetry with Higgs fields on the two branes (see e.g.~\cite{Cacciapaglia:2005pa,Csaki:2005vy}). The above boundary conditions are then obtained in the limit of their VEVs going to infinity.
In unitary gauge, $\xi \rightarrow \infty$, the bulk equation of motion for $A_5$ gives
\begin{equation}
\label{eq:bulk}
\partial_5 \Bigl( a(z)^{-1}\partial_5 \bigl( a(z) A_5 \bigr)\Bigr) \,  = \, 0 \, .
\end{equation}
Notice that this equation is consistent with the boundary conditions and there is thus one massless mode from $A_5$. Its other Kaluza-Klein modes are all eaten by $A_\mu$. In particular, there is no massless mode from $A_\mu$, consistent with the fact that the gauge symmetry is broken.
As usual, the $A_5$ massless mode can be parameterized as
\be
\label{eq:A50}
A_5(x,z) \, = \, h(z) \,\phi(x) \, ,
\ee
where $h(z)$ is its profile along the extra dimension. From Eqs.~\eqref{e:bcs} and \eqref{eq:bulk}, we then see that
$h(z)= \mathcal{N} a(z)^{-1}$. Demanding canonically normalized kinetic terms for $\phi(x)$, the normalization constant $\mathcal{N}$ of the wavefunction is determined by
\be
\label{eq:normalization}
\frac{\mathcal{N}^2}{g_5^2}\int_{z_\UV}^{z_\IR} \frac{dz}{a(z)} \, = \, 1 \, .
\ee
For $kL\gg1$, this gives $\mathcal{N} \simeq  g_4 \sqrt{2 k L} \,e^{-kL}$, where we define the dimensionless coupling $g_4 \equiv g_5/\sqrt{L}$. Altogether, the wavefunction of the massless mode then reads
\begin{align}
\label{A5Wavefunction}
h(z) \, \simeq \, g_4 \sqrt{2 k L} \, e^{-kL} kz  \, .
\end{align}
The wavefunction is thus peaked towards the IR brane (see \Fig{fig:warped} for a sketch of the wavefunction profile in the extra dimension). Furthermore, the fact that $\mathcal{N} \rightarrow 0$ for $z_\IR \rightarrow \infty$ shows that the $A_5$ massless mode is indeed localized in the IR.

Performing a 5D gauge transformation, $A_M(x,z)\rightarrow A_M(x,z)+ \partial_M \alpha(x,z)$, we see that the boundary conditions in \Eq{e:bcs} and the bulk equation of motion in Eq.~\eqref{eq:bulk} remain invariant only for the subset of transformations
\be
\label{RemnantSymmetry}
\alpha = B \, z^2 \, + \, C
\ee
with $B$ and $C$ being independent of $x$ and $z$. The remaining symmetry in 4D is thus global, again consistent with the fact that the gauge symmetry is broken. Under this remnant symmetry, the massless mode transforms as
\be
\label{PhiShift}
\phi \, \rightarrow \, \phi \, + \, \frac{2 B}{\mathcal{N} k} \, .
\ee

At this point, the relaxion is thus an exact Nambu-Goldstone boson which non-linearly realizes a remnant global $U(1)$. By virtue of the 5D gauge invariance, no 5D local, higher-dimensional operators can break this shift symmetry (see \cite{Choi:2003wr} for a detailed discussion).
A potential for the relaxion could be generated by non-local effects in the presence of bulk states which are charged under the $U(1)$ but we assume such states to be absent from the theory.\footnote{Alternatively, for example for bulk fermions charged under the $U(1)$ it is sufficient if their masses are somewhat larger than the AdS scale in which case any perturbative contribution to the potential is highly suppressed (see e.g.~\cite{Contino:2003ve,Pilo:2003gu}).} Instead we introduce anomalous couplings of the relaxion to confining non-abelian gauge groups. A potential then arises from instantons, similar to what happens for the axion in QCD.
We localize these anomalous couplings in the bulk or on the UV brane. In what follows, we show that these possibilities, thanks to the warp factor, can naturally explain the required hierarchy between the decay constants in the relaxion potential.

\subsection{Anomalous couplings from the bulk}
\label{ss:bulkfermions}

Let us add a non-abelian gauge group in the bulk, whose field strength and coupling we denote respectively as  $\bm{G}_{NP}$ and $g^c_5$. We choose boundary conditions for the gauge field such that the 4D gauge symmetry remains unbroken on the branes. Its tower of Kaluza-Klein modes then contains one massless mode which is the 4D gauge boson. We next introduce a Chern-Simons coupling of the $U(1)$ gauge field to this gauge group. Including the kinetic term, the action reads
\be
\label{eq:cs}
S_{\rm{5D}} \, \supset \, \int d^4x \, dz\, \left(  \sqrt{g} \frac{-1}{2 (g^c_5)^2} \mathrm{Tr} \left[  \bm{G}_{MN} \bm{G}^{MN} \right]\, + \, \frac{c_\B}{16 \pi^2} \epsilon^{MNPQR} A_M \mathrm{Tr} \left[  \bm{G}_{NP} \bm{G}_{QR} \right] \right) \, ,
\ee
where $c_\B$ is a dimensionless constant and the normalization is chosen for later convenience.\footnote{Note that a factor of $2$ arises from the normalization $\mathrm{Tr}[T^a T^b]= \frac{1}{2}\delta_{a,b}$ of the generators of the non-abelian gauge group.} Under a $U(1)$ gauge transformation $A_M(x,z)\rightarrow A_M(x,z)+ \partial_M \alpha(x,z)$, the action transforms as
\be
\label{AnomalousTransformation}
S_{\rm 5D} \, \rightarrow \, S_{\rm 5D} \, -  \int d^4x \, dz \, \alpha(x,z)  \frac{c_\B}{16 \pi^2} \epsilon^{\mu \nu \rho \sigma} \mathrm{Tr} \left[ \bm{G}_{ \mu \nu} \bm{G}_{\rho \sigma} \right] \, \bigl( \delta(z-z_\UV) \, - \, \delta(z-z_\IR) \bigr)  \, .
\ee
The Chern-Simons term thus induces an anomaly for the $U(1)$ symmetry on the branes. This is not a problem, however, since the symmetry is only global on the branes and there are thus no gauge anomalies.

In the 4D effective theory, this gives rise to an anomalous coupling for $\phi$. Let us restrict ourselves to the massless mode of the non-abelian gauge field, whose field strength we denote as $G_{\mu \nu}$.
Integrating over the extra dimension, \Eq{eq:cs} then in particular gives
\be
\label{eq:cseffective}
S_{\rm{4D}} \, \supset \,  \int d^4x \, \left(\frac{-1}{2 (g_4^c)^2} \mathrm{Tr} \left[ G_{\mu \nu} G^{\mu \nu} \right] \, + \, \frac{\phi(x)}{16 \pi^2 f_\B} \epsilon^{\mu \nu \rho \sigma} \mathrm{Tr} \left[ G_{\mu \nu} G_{\rho \sigma} \right] \right) \, ,
\ee
where $g_4^c = g_5^c /\sqrt{L}$ is the gauge coupling of the massless mode. The decay constant is given by \cite{Choi:2003wr,Flacke:2006ad}
\begin{align}
\label{eq:fB}
f_\B \equiv \left[c_\B \int_{z_\UV}^{z_\IR} dz \, h(z)\right]^{-1}  = \, \frac{\mathcal{N}}{c_\B g_5^2} \, \simeq \, \frac{2k\,e^{-k L}}{ c_\B\, g_4 \sqrt{2kL}}
\end{align}
which is of order the IR scale $\Lambda_\IR$ and thus warped-down. From Eqs.~\eqref{PhiShift}, \eqref{AnomalousTransformation} and \eqref{eq:cseffective}, we see that $\phi$ reproduces the anomaly under a transformation $\alpha = B z^2$. In Appendix~\ref{AppendixA}, we briefly review how Chern-Simons terms can arise from charged bulk fermions. As we also discuss there, any perturbative contribution from such a fermion to the potential for $A_5$ can be sufficiently suppressed. Nevertheless, in the remainder of this paper we will never assume any charged bulk states and will instead include the Chern-Simons terms directly into our effective 5D theory.

Note that \Eq{eq:cs} also yields couplings of $\phi$ to the higher Kaluza-Klein modes of the non-abelian gauge field. As \Eq{eq:cseffective} for the massless mode, these couplings are total derivatives (see e.g.~\Ref{Grzadkowski:2007xm}) and therefore do not contribute perturbatively to the potential for $\phi$. We will later assume that the non-abelian gauge group confines in order to generate a non-perturbative potential for $\phi$. But we will choose the confinement scale below the IR scale and thus below the Kaluza-Klein masses. The Kaluza-Klein modes of the non-abelian gauge group therefore do not contribute non-perturbatively to the potential either.

\subsection{Anomalous couplings from the UV brane}
\label{ss:uvfermions}
We now discuss how a decay constant which is much larger than $\Lambda_\IR$ can be obtained. To this end, we consider an effective anomalous coupling of $A_5$ which is localized on the UV brane \cite{Flacke:2006ad},
\be
\label{eq:intlocalizedUV}
S_{\rm 5D} \, \supset \, \int d^4x \, dz \, \delta(z - z_\UV)  \frac{c_\UV}{16 \pi^2}\frac{A_5}{k}~ \epsilon^{\mu \nu \rho \sigma} \, \mathrm{Tr} \left[ \bm{G}_{\mu \nu} \bm{G}_{\rho \sigma} \right]  ,
\ee
where $c_\UV $ is a dimensionless constant and $\bm{G}_{MN}$ is the field strength of a non-abelian gauge field in the bulk. As we outline in Appendix~\ref{Appendix1}, this interaction can for example arise as an effective coupling from a Chern-Simons term in a two-throat geometry. Under a $U(1)$ gauge transformation, the action transforms similar to Eq.~\eqref{AnomalousTransformation} but restricted to the UV brane and with $\partial_5 \alpha(x,z)$ instead of $\alpha(x,z)$.
Let us again restrict ourselves to the massless mode of the gauge field, whose field strength we denote as $G_{\mu \nu}$. Using the wavefunction of the massless mode of $A_5$ from Eqs.~\eqref{eq:A50} and \eqref{A5Wavefunction}, this gives
\be
\label{eq:intUV}
S_{\rm 4D} \, = \, \int d^4x \, \frac{1}{16 \pi^2}\, \frac{ \phi(x)}{f_\UV} \, \epsilon^{\mu \nu \rho \sigma} \, \mathrm{Tr} \left[ G_{\mu \nu} G_{\rho \sigma} \right]
\ee
with decay constant given by \cite{Flacke:2006ad}
\begin{align}
\label{eq:fUV}
f_\UV \equiv   \frac{k\,}{c_\UV \, h(z_\UV)} \, \simeq \, \frac{  k\, e^{kL}}{ c_\UV \, g_4\,\sqrt{2kL}}
\end{align}
or $f_\UV \sim M_\PL^2/\Lambda_\IR$.  We see that a warped-up decay constant, much larger than the cutoff, appears naturally in this case.
This large decay constant can be intuitively understood as being of order the natural scale $M_\PL$ on the UV brane times an inverse suppression factor from the wavefunction overlap of $A_5$ with the UV brane.

Note that super-Planckian decay constants may be constrained by the weak gravity conjecture in theories of quantum gravity \cite{ArkaniHamed:2006dz} (see also \cite{Hebecker:2015zss, Ibanez:2015fcv, Heidenreich:2015wga, Heidenreich:2015nta}). Given that the relaxion is an axion-like field, the conjecture  necessarily  restricts its field excursion ($\Delta \phi \sim \Lambda/g'$) to be sub-Planckian, setting a lower bound on the coupling $g'$ in the potential (\ref{eq:VRelaxion1a}). The weak gravity conjecture is then at odds with any relaxion model with trans-Planckian field excursions, including our proposal.
On the other hand, there are known loopholes to the conjecture \cite{Rudelius:2015xta, Brown:2015iha, Hebecker:2015rya, Saraswat:2016eaz, Ibanez:2017vfl}.
 For instance, the application  of the conjecture to effective field theories may result in a much weaker bound on the coupling $g'$ \cite{Saraswat:2016eaz}. Furthermore, in \cite{Ibanez:2017vfl}, a better understanding of the conclusions of \cite{Saraswat:2016eaz} is achieved by considering a string theory embedding. There it is shown that if a clockwork model is successfully embedded in string theory, one may in principle obtain a large cutoff, avoiding the naive bound from the weak gravity conjecture, as long as the number of sites in the construction is large.

We conclude that two hierarchically different decay constants can be obtained, depending on the localization of the anomalous interactions in the warped space. For the relaxion, we then identify $F= f_\UV \approx M_\PL^2/\Lambda_{\IR}$ and $f=f_\B \approx \Lambda_\IR$.
Note that as the ratio $F/f$ is proportional to the warp factor, the potential in \Eq{eq:VRelaxion2a} does not respect a discrete shift symmetry since, in general,  $F/f$  is a non-integer number. This is a consequence of the non-local nature of the residual symmetry transformation $\alpha = B z^2 + C$ in \Eq{RemnantSymmetry} which explicitly depends on the localization.
In the following, we  build an explicit model that makes use of this toolkit to generate a phenomenologically viable potential in the form of \Eq{eq:VRelaxion2a}.

\section{Generating the relaxion potential}
\label{sec:RelaxionPotentialFromA5}

\subsection{General setup}
\label{GeneralSetup}
Let us next discuss the relaxion parameters in more detail and how they can be understood in terms of our UV model. Provided that electroweak symmetry remains unbroken in the confinement phase transition which generates the periodic potentials in \Eq{eq:VRelaxion2a},
$\Lambda_{F,f}(H)$ both depend quadratically on the Higgs (plus generically higher even powers of the Higgs which are, however, not important in the following).\footnote{As proposed in \cite{Graham:2015cka}, one can also use the QCD axion as the relaxion. The last term in Eq.~\eqref{eq:VRelaxion1a} is then the usual QCD axion potential which depends linearly on the Higgs (see e.g.~\cite{diCortona:2015ldu}). However, barring additional model building, this spoils the axion solution to the strong $CP$ problem. See also \cite{Nelson:2017cfv, Jeong:2017gdy, Davidi:2017gir} for solutions to the strong $CP$ problem in the context of the relaxion.} We can then parametrize
\be
\Lambda_{F,f}^4(H) \, = \, \Lambda_{F,f}^4 \, \left(1 \, + \, \frac{H^2}{M_{F,f}^2} \right) \, ,
\ee
where $\Lambda_{F,f}$ and $M_{F,f}$ can be understood as the scales where the periodic terms and higher-dimensional couplings to the Higgs  are generated, respectively. The potential in \Eq{eq:VRelaxion2a} then reads
\begin{equation}
\label{eq:Vrelaxion2b}
V(\phi, H) \, = \, - \Lambda^2 H^2 \, + \, \lambda \, H^4 \, + \, \Lambda_\cF^4 \left( 1 \, + \, \frac{H^2}{M_F^2} \right)  \cos{\left(\frac{\phi}{F}\right)} + \, \Lambda_\cf^4 \, \left(1 \, + \, \frac{H^2}{M_f^2} \right) \cos\left(\frac{\phi}{f}\right) \, .
\end{equation}
For simplicity, we have dropped terms which may be generated at higher loop-order. We will discuss these terms later in \Sec{sec:constraints}. Assuming that $\phi$ is in the linear regime of the low-frequency cosine, $\phi \sim \pi F/2 \; \text{mod} \; 2 \pi$, we can expand it for $\phi-\pi F/2 \lesssim F$. After the redefinition $\phi-\pi F/2 \rightarrow \phi$, this gives the linear part of the relaxion potential in \Eq{eq:VRelaxion1a} with the identifications
\be \label{e:CouplingIdentification}
g \, = \, \frac{ \Lambda_\cF^4}{F  \Lambda^3}, \, \quad \qquad  g' \, = \, \frac{\Lambda_\cF^4}{F M_F^2 \Lambda}
\ee
up to factors of order one.

The last term in \Eq{eq:Vrelaxion2b} stops the relaxion once the Higgs VEV has reached the electroweak scale. For this to work, we need to ensure that $M_f \lesssim v_\ew $, otherwise the Higgs-independent barrier proportional to $\cos (\phi/f)$ would  stop the relaxion already before the Higgs VEV has obtained the right value. Note also that the Higgs-independent barrier receives corrections from closing the Higgs loop in the Higgs-dependent one and will thus generically be present. We discuss radiative corrections to the potential in more detail in \Sec{sec:constraints}. But to get
a sense of the scales involved, we already note here that radiative stability of the potential demands that $\Lambda_f^2 \lesssim 4 \pi \, v_\ew M_f$ and $\Lambda_F  \lesssim 4\pi M_F$.

To obtain $M_f \lesssim v_\ew $ requires that the higher-dimensional coupling of the Higgs to the periodic potential is generated near the electroweak scale. In the next section, we make use of a construction from \Ref{Graham:2015cka} which introduces light fermions for this purpose.
The drawback of this scenario is of course a coincidence problem: one has to assume new particles at a scale which is dynamically generated by the relaxation mechanism and is thus determined by a priori completely unrelated parameters. One way around this problem is the double-scanner mechanism of Ref.~\cite{Espinosa:2015eda}. To this end, one introduces another axion-like field which dynamically cancels off the Higgs-independent barrier in \Eq{eq:Vrelaxion2b}. This allows the relaxation mechanism to work even for $M_f \gg v_\ew $.\footnote{Another proposal for the relaxion that does not require new physics close to the electroweak scale is the particle-production mechanism of \Ref{Hook:2016mqo}.}
We discuss a UV completion of this scenario in \Sec{sec:doublescanner}.

\subsection{A warped model}
\label{ss:toymodel}

We now build a simple explicit model that successfully generates the needed terms in the Higgs-relaxion potential at a phenomenologically viable scale, making use of the results of \Sec{sec:HierarchicalDecayConstants}. We assume that the Higgs is localized on or near the IR brane, so that its mass is warped down to the IR scale (see \Fig{fig:warped}). We note that it may also be possible to implement the relaxation mechanism in a model where the Higgs is instead localized on the UV brane. As usual, the relaxion can only protect the Higgs up to some cutoff significantly below the Planck scale. Such a model would therefore require a UV completion above this cutoff on the UV brane. We leave a study of this possibility to future work.  As we find later, the highest IR scale that we can achieve in our implementation of the relaxation mechanism (while still solving the hierarchy problem) is below the GUT scale. If the remaining Standard Model fields are then also localized on the IR brane, higher-dimensional operators violating baryon number lead to too fast proton decay \cite{Gherghetta:2000qt}. In order to suppress these operators, we assume that the Standard Model instead lives in the bulk. As usual, the light quarks are localized towards the UV brane, while the top-bottom doublet and the right-handed top live nearer to the IR brane. This has the added advantage that the hierarchy of Yukawa couplings can then be generated from the warping too. The IR scale in our model can be high enough, on the other hand, to ensure that oblique corrections and flavour- and $CP$-violating processes are sufficiently suppressed without imposing custodial or flavour symmetries.

We identify the relaxion with the 5th component of a $U(1)$ gauge field in the bulk. In order to generate a potential for this field, we add two non-abelian gauge groups $\mathcal{G}_f$ and $\mathcal{G}_F$ which also live in the bulk.
We assume that these gauge groups confine at the scales $\Lambda_{\mathcal{G}_f}$ and $\Lambda_{\mathcal{G}_F}$, respectively. In order to ensure that confinement can be discussed using only the zero-modes of the bulk gauge fields, we take $\Lambda_{\mathcal{G}_f}$ and $\Lambda_{\mathcal{G}_F}$ to be below the IR scale. This can always be arranged by choosing the 5D gauge couplings and ranks of the gauge groups appropriately.

We assume anomalous couplings of the relaxion $\phi$ to the field strengths $G^{f}_{\mu\nu}$ and $G^{F}_{\mu\nu}$ of the massless 4D gauge fields corresponding to $\mathcal{G}_f$ and $\mathcal{G}_F$, respectively:
\be
\label{e:f}
S_{\rm 4D} \, \supset\, \int d^4x \, \frac{\phi(x)}{16 \pi^2} \, \epsilon^{\mu \nu \rho \sigma} \left(\frac{1}{F}  \mathrm{Tr} \left[G^{F}_{\mu \nu} G^{F}_{\rho \sigma}\right] \, + \, \frac{1}{f}  \mathrm{Tr} \left[G^{f}_{\mu \nu} G^{f}_{\rho \sigma}  \right]\right) \, .
\ee
As we have discussed in \Sec{sec:HierarchicalDecayConstants}, these can arise from a Chern-Simons coupling in the bulk and an effective anomalous coupling of $A_5$ on the UV brane.
But for now, we only assume that $F \gg f$ and postpone a concrete choice for the decay constants to \Sec{sec:constraints}.

On the IR brane, we add a pair of chiral fermions $\chi$ and $\chi^c$ in the fundamental and antifundamental representation of $\mathcal{G}_F$, respectively. These fermions transform under a chiral symmetry which we assume to be broken only by a Dirac mass $m_\chi$. This allows for the terms in the action
\be
\label{e:mir0}
S_{\rm 5D} \, \supset \, \int d^4x \, dz \, \sqrt{-g_\IR} \, \delta(z-z_\IR) \, m_\chi \left( 1 + \frac{H^2}{M_\PL^2} \right) \chi\, \chi^c \, + \, \text{h.c.}\, ,
\ee
where $g_\IR$ is the induced metric determinant on the IR brane. We have included a higher-dimensional coupling to the Higgs which is generically present and which we expect to be suppressed by a scale near the Planck scale. Note that we will use the symbol $H$ for both the $SU(2)$-doublet Higgs field, writing the singlet combination $|H|^2$ as $H^2$ for simplicity, and its VEV. It will be clear from context which one is meant.
For simplicity, we also ignore any numerical prefactors for now and set $k=M_\PL$. Similarly, we assume that all parameters are real. We will reinstate prefactors and phases later on. Performing the integral over the extra dimension and canonically normalizing the fields gives
\be
\label{e:mir}
S_{\rm 4D} \, \supset \, \int d^4x \, \,m_\chi \left( 1 + \frac{H^2}{\Lambda_\IR^2} \right) \chi\, \chi^c \, + \, \text{h.c.}\, ,
\ee
where we have redefined $ e^{-kL} m_\chi \rightarrow m_\chi$, $ e^{-kL} H \rightarrow H$, $ e^{-3kL/2} \chi \rightarrow \chi$ and similarly for $\chi^c$. Note in particular that $m_\chi \lesssim \Lambda_\IR$ after the redefinition. Let us next perform the field redefinition
\be
\label{ChiralRotationChi}
\chi \, \rightarrow  \, e^{i \phi / F} \chi \, ,
\ee
while $\chi^c$ is left invariant. Due to the non-trivial transformation of the path integral measure, this chiral rotation removes the coupling of $\phi$ to $\mathrm{Tr} \left[G^{F}_{\mu \nu} G^{F}_{\rho \sigma}\right]$ in \Eq{e:f} and transforms \Eq{e:mir} to
\be
\label{S4ChiF}
S_{\rm 4D} \, \rightarrow \, S_{\rm 4D} \supset \int d^4x \, m_\chi \left( 1 \, + \, \frac{H^2}{\Lambda^2_\IR}\right) e^{i \phi / F} \chi \chi^c \, + \, \text{h.c.}\, .
\ee
If $m_\chi$ is below the confinement scale of $\mathcal{G}_F$ (which in turn is below $\Lambda_\IR$), this term contributes to the Higgs-relaxion potential after confinement. Parametrizing\footnote{This is thus our definition of the scale $\Lambda_{\mathcal{G}_F}$.
}  $\langle \chi \chi^c \rangle = \Lambda_{\mathcal{G}_F}^3$, this gives
\be
\label{e:potF}
V(\phi,H) \, \supset \, m_\chi \, \Lambda_{\mathcal{G}_F}^3 \left( 1 \, + \, \frac{H^2}{\Lambda_\IR^2}\right)  \, \cos\left(\frac{\phi}{F} \right)  \,.
\ee
This has the same form as the potential with period $F$ in \Eq{eq:Vrelaxion2b}, including the coupling to the Higgs. We can then make the identifications
\be
\label{e:dictionary}
 \Lambda_\cF^4 \, = \, m_\chi \, \Lambda_{\mathcal{G}_F}^3 \, , \qquad \   M_F^2 = \Lambda_\IR^2 \,.
\ee

\begin{table}
\begin{center}
\renewcommand{\arraystretch}{1.2}
\begin{tabular}{| c || c | c | c | c | c | c |}
\hline
 & $\chi$ & $\chi^c$ & $N$ & $N^c$ & $ L$  & $L^c$ \tabularnewline
\hline
\hline
$\mathcal{G}_F$ &  $ \square$ & $\bar{\square}$ & \bf{--} &  \bf{--} &  \bf{--} & \bf{--} \tabularnewline
\hline
$\mathcal{G}_f$   & \bf{--} & \bf{--} &  $\square$ & $\bar{\square}$ &  $\square$ & $\bar{\square}$ \tabularnewline
\hline
$SU(2)_L$   & \bf{--} & \bf{--} & \bf{--} & \bf{--}  &  $\square$ & $\square $ \tabularnewline
\hline
$U(1)_Y$   & \bf{--} & \bf{--} & \bf{--} &  \bf{--}  &  $-\frac{1}{2}$ & $+\frac{1}{2}$\tabularnewline
\hline
\end{tabular}
\caption{\label{t:matter}\emph{Matter content on the IR brane with gauge representations for the model with a barrier at the electroweak scale.  }}
\end{center}
\end{table}

Next we need to generate the potentials with smaller period $f$. To this end, we use a construction from \Ref{Graham:2015cka} and add fermions $L$ and $N$ on the IR brane with the same Standard Model charges as the lepton doublet and the right-handed neutrino, respectively. In addition, these fermions are in the fundamental representation of the gauge group $\mathcal{G}_f$. We also include fermions $L^c$ and $N^c$ in the conjugate representations. Together they allow for the terms in the action
\be
\label{e:mirt0}
S_{\rm 5D} \, \supset \, \int d^4x \, dz \, \sqrt{-g_\IR} \, \delta(z-z_\IR) \, \left( \, m_L \, L L^c \, + \, m_N \, N N^c \, + \, y \, H L N^c \, + \, \tilde{y} \, H^{\dagger} L^c N \, \right)\, + \, \text{h.c.} \, .
\ee
Notice that we have not included a higher-dimensional coupling to the Higgs. It could be present but will be subdominant as we will see momentarily.
Performing the integral over the extra dimension and canonically normalizing the fields gives
\be
\label{e:mirt}
S_{\rm 4D} \, \supset \, \int d^4x \, \left( \, m_L \, L L^c \, + \, m_N \, N N^c \, + \, y \, H L N^c \, + \, \tilde{y} \, H^\dagger L^c N \, \right)\, + \, \text{h.c.} \, ,
\ee
where we have redefined $ e^{-kL} m_L \rightarrow m_L$, $ e^{-kL} H \rightarrow H$, $ e^{-3kL/2} L \rightarrow L$ and similarly for $m_N$, $N$ and the conjugated fields. Note in particular that $m_L, m_N \lesssim \Lambda_\IR$ after the redefinition.
Assuming that $m_N \ll m_L$ and restricting to a region in field space where the Higgs VEV satisfies $y \tilde{y} H^2 \ll m_L^2 $, we can integrate out $L$ and $L^c$. This gives
\begin{align}
\label{e:mirtN}
S_{\rm 4D} \, \supset \, \int d^4x \, \left(  \, m_N \,  - \frac{y\tilde{y}\,H^2}{m_L} \,\right)\,N N^c \,\, + \, \text{h.c.} \, .
\end{align}
We can then perform the chiral rotation
\be
N \, \rightarrow \, e^{i \phi/f} N \, ,
\ee
while $N^c$ is left invariant. This removes the coupling of $\phi$ to $\mathrm{Tr} \left[G^{f}_{\mu \nu} G^{f}_{\rho \sigma}\right]$ in \Eq{e:f} and transforms Eq.~\eqref{e:mirtN} to
\begin{align}
S_{\rm 4D} \, \rightarrow \, S_{\rm 4D} \, \supset \, \int d^4x \, \left(  \, m_N \,  - \frac{y\tilde{y}\,H^2}{m_L} \,\right)\, e^{i \phi /f} \,N N^c \,\, + \, \text{h.c.} \, .
\end{align}
Provided that $m_N$ is below the confinement scale of $\mathcal{G}_f$, this term contributes to the Higgs-relaxion potential after confinement. Parametrizing $\langle N N^c \rangle = \Lambda_{\mathcal{G}_f}^3$, this gives
\be
\label{e:potf}
V(\phi,H) \, \supset \, m_N \, \Lambda_{\mathcal{G}_f}^3  \left(1  \, - \, \frac{y \tilde{y} \, H^2}{m_N m_L}\, \right)   \, \cos\left(\frac{\phi}{f} \right)  \,.
\ee
This has the form of the potential with period $f$ in \Eq{eq:Vrelaxion2b}, including the coupling to the Higgs. We can then make the identifications
\be
\label{e:dictionary2}
\Lambda_\cf^4 \, = \, m_N \, \Lambda_{\mathcal{G}_f}^3 \, , \qquad \  M_f^2 = \frac{m_N m_L}{y\tilde{y}} \,.
\ee
For sufficiently small $m_N$ and $m_L$, this allows for $M_f \lesssim v_\ew$ as required in a technically natural way. Notice that if we had instead relied on the higher-dimensional operator in \Eq{e:mir0} to generate the barrier, we would have obtained $M_f \sim \Lambda_\IR \gg v_\ew$. We discuss constraints on the parameters of this construction in more detail in \Sec{sec:constraints}. A summary of the matter content on the IR brane is given in \Tab{t:matter}.

We next reinstate the numerical prefactors and the phases of the parameters which we have ignored so far. Let us denote the prefactor of the Higgs coupling in \Eq{e:mir0} as $c_{\chi H}$. We absorb possible phases in the fermionic condensates $\langle \chi \chi^c \rangle$ and $\langle N N^c \rangle$ and any (relaxion-independent) $\Theta$-terms for $\mathcal{G}_F$ and $\mathcal{G}_f$
into the mass parameters $m_\chi$ and $m_N, m_L$, respectively. Redoing the derivation above then gives
\begin{multline}
\label{PotentialWithPhases}
V(\phi,H) \, \supset \, 2 |m_\chi | \, \Lambda_{\mathcal{G}_F}^3 \left[ \cos\left(\frac{\phi}{F} + b_\chi\right) \, + \, |c_{\chi H} | \, \frac{H^2}{\Lambda_\IR^2}  \, \cos\left(\frac{\phi}{F} + b_{\chi H}\right) \right] \\
+ \, 2 |m_N| \, \Lambda_{\mathcal{G}_f}^3  \left[\cos\left(\frac{\phi}{f} + b_N \right)  \, - \, \frac{|y \tilde{y}| \, H^2}{|m_N m_L|}\,\cos\left(\frac{\phi}{f} + b_{NH}\right) \right]  \, ,
\end{multline}
where the complex phases are given by $b_\chi = \arg(m_\chi)$, $b_{\chi H} = \arg(m_\chi c_{\chi H})$, $b_N = \arg(m_N)$ and $b_{N H} = \arg(y \tilde{y} / m_L)$. Note that this does generically not match the form of the potential in \Eq{eq:Vrelaxion2b}. Nevertheless the relaxation mechanism can still work. Indeed expanding the first two terms in the linear part of the cosines again gives the sliding term for the relaxion and its linear coupling to the Higgs. In order to ensure that these terms have the same sign as required, we need to demand that $b_\chi \sim b_{\chi H}$. As before, the Higgs-independent barrier in the third term should be too small to stop the relaxion by itself. It is then negligible for the dynamics and the phase $b_N$ has no consequences. The phase $b_{N H}$ in the Higgs-dependent barrier in the fourth term, on the other hand, slightly shifts the minimum where the relaxion eventually stops but has no other consequences either.

To ensure that our calculation of the potentials is applicable, the masses of the fermion pairs $\chi, \chi^c$ and $N, N^c$ need to be below their respective condensation scales. This means that the chiral symmetries under which these fermion pairs transform are only weakly broken at the confinement scales. We then expect corresponding pseudo-Nambu-Goldstone bosons in the spectrum of composite states. As we discuss in Appendix \ref{AppendixB}, their contribution to the potential factorizes from the remaining potential and they can be trivially integrated out if the spectrum of fermions is doubled.

\section{Conditions for successful relaxation}
\label{sec:constraints}
We now discuss various conditions that need to be fulfilled for the relaxation mechanism to be viable. In \Sec{sec:GeneralConditions}, we derive general conditions on the parameters in the relaxion potential in \Eq{eq:Vrelaxion2b}. In \Sec{sec:ConditionsFromEWBarrier}, we then discuss additional conditions that arise in our warped model with a barrier at the electroweak scale.

\subsection{General conditions}
\label{sec:GeneralConditions}
We begin our discussion of the evolution of the Higgs and relaxion with the Higgs mass-squared being positive and of order $\Lambda^2$. In order to allow the relaxion to subsequently turn the Higgs mass tachyonic, its average VEV $\tilde{\phi}$ during this stage of the evolution needs to satisfy
\be
\label{PhiChange}
\cos \Bigl(\frac{\tilde{\phi}}{F}\Bigr)  \, \gtrsim \, \frac{\Lambda^2 \, M_F^2}{\Lambda_F^4} \, .
\ee
Since the left-hand side is bounded by 1, this in particular implies the condition
\be
\label{ExpansionConstraint}
\Lambda_F^2 \, \gtrsim \, \Lambda \, M_F \, .
\ee

The relaxion stops rolling down its potential when the derivatives of the periodic terms balance each other. We will find below that $M_F \gg v_\ew$ and the term proportional to $\cos (\phi/F)$ is thus dominated by the Higgs-independent part. On the other hand, the term proportional to $\cos (\phi/f)$ needs to be dominated by the Higgs-dependent part as discussed in Sec.~\ref{sec:RelaxionPotentialFromA5}.
The relaxion then stops once the Higgs VEV becomes
\be
\label{EWscalerelation}
H^2 \, \approx \, M_f^2 \, \frac{f}{F} \frac{\Lambda_\cF^4}{\Lambda_\cf^4} \, ,
\ee
where we have set $\sin(\tilde{\phi}/F)\sim 1$. This is a good approximation as long as $\cos(\tilde{\phi}/F)$ is not very close to its extrema. The parameters need to be chosen such that the combination on the right-hand side gives the electroweak scale $v_\ew$. In the following, we will use this relation to trade $\Lambda_\cf$ for $v_\ew$.

Notice that the Higgs-dependent barrier $H^2 \cos(\phi/f)$ in the potential contributes to the Higgs mass. Imposing that this contribution be less than the electroweak scale  (see e.g.~Ref.~\cite{Flacke:2016szy})\footnote{This constraint can be slightly relaxed if one includes the barrier term in the scanning of the Higgs mass \cite{Choi:2016luu}. One then still needs to impose that $\Lambda_f^2 \lesssim 4 \pi M_f v_\ew$ to ensure that loop corrections to the potential are small. This gives a similar condition as \Eq{eq:barrierValue} but with an additional factor $\sqrt{4\pi}$ on the right-hand side.
} gives the constraint $\Lambda_f^2 \lesssim M_f v_\ew$ which using \Eq{EWscalerelation} leads to
\begin{equation}
\label{eq:barrierValue}
\Lambda_F \, \lesssim \, v_{\ew} \left(\frac{F}{f}\right)^{1/4}  .
\end{equation}
Together with \Eq{ExpansionConstraint}, this gives a constraint on the cutoff in our model as we discuss in \Sec{sec:ConditionsFromEWBarrier}. In order to ensure that the Higgs mass is scanned with sufficient precision,  we need to demand that the change of the Higgs-dependent term proportional to $\cos(\phi/F)$ over one period of the barrier, $\delta \phi \sim f$, is less than the electroweak scale. This gives the constraint $\Lambda_F \lesssim (M_F \, v_\ew)^{1/2} (F/f)^{1/4} $ which is weaker than \Eq{eq:barrierValue}.

Furthermore, there are several requirements on the inflation sector for the relaxation mechanism to be viable. If the relaxion is not the inflaton, its energy density should be subdominant compared to the inflaton. The energy density in the minimum where the relaxion eventually settles needs to be (close to) zero. This requires an additional constant contribution that is added to the potential and chosen such that the energy density at the minimum (nearly) vanishes. The tuning that is necessary to achieve this is just a manifestation of the cosmological constant problem. The contribution of the relaxion to the energy density relevant for inflation is then determined by how much it changes during its evolution. Using \Eq{PhiChange} in the potential of \Eq{eq:Vrelaxion2b} gives the condition
\be
\label{LowerLimitHI}
H_I \, \gtrsim \, \frac{M_F \Lambda}{M_\PL} \, ,
\ee
where $H_I$ is the Hubble rate during inflation. In addition, to ensure that our classical analysis of the field evolution is applicable, quantum fluctuations of the relaxion while it roles down the potential should be sufficiently small. Over one Hubble time, the relaxion changes classically by $(\delta \phi)_{\rm class.} \sim H_I^{-2} dV/d\phi$. Its quantum fluctuations, on the other hand, are $(\delta \phi)_{\rm quant.} \sim H_I$. This leads to the condition
\be
\label{UpperLimitHI}
H_I \, \lesssim  \, \frac{\Lambda_\cF^{4/3}}{F^{1/3}} \, .
\ee
Combining the last two inequalities, we get
\be \label{InflationConstraint}
\Lambda_\cF^2 \, \gtrsim \, \sqrt{F} \left(\frac{ M_F\, \Lambda}{ M_\PL} \right)^{3/2} \, .
\ee
Finally, the number of e-folds of inflation must be sufficiently large to ensure that the relaxion scans the required field range. Denoting the latter by $\Delta\phi$, this leads to the condition $\mathcal{N}_e (\delta \phi)_{\rm class.} \gtrsim \Delta \phi$. Provided that the relaxion is in the linear part of $\cos (\phi/F)$, using \Eq{PhiChange} this gives
\begin{equation}
\label{efoldsLimit}
\mathcal{N}_e \, \gtrsim \, \left(\frac{H_I F  M_F  \Lambda}{\Lambda_F^4}\right)^2.
\end{equation}
The resulting required number of e-folds can be very large. We will not specify the inflation sector but will simply assume that it can be arranged to fulfill the conditions in Eqs.~\eqref{LowerLimitHI}, \eqref{UpperLimitHI} and \eqref{efoldsLimit}. Possible complications in achieving this are discussed e.g.~in \Ref{DiChiara:2015euo}. Note also that the above conditions are somewhat alleviated if the effect of the time evolution of the Hubble rate during inflation is taken into accout \cite{Patil:2015oxa}.

We also need to ensure that the potential is radiatively stable. The potential is an effective theory with a cutoff determined by the confinement scales $\Lambda_{\mathcal{G}_f}$ and $\Lambda_{\mathcal{G}_F}$ of the gauge groups that give rise to the periodic terms (assuming they are smaller than the cutoffs associated with generating the $H^2$-terms in the potential). In the region of the potential where the Higgs mass parameter\footnote{Note that the Higgs mass parameter has an additional contribution from the $\cos(\phi/f)$-term. Since it is subdominant except in a small region of $\phi$, we define \Eq{HiggsMassParameter} without this contribution.}
\be
\label{HiggsMassParameter}
m_H^2(\phi) \,\equiv  \frac{\Lambda_F^4}{M_F^2} \, \cos \left(\frac{\phi}{F}\right) \, - \,\Lambda^2
\ee
is smaller than these cutoffs, the Higgs can give important corrections to the potential. From the one-loop effective potential, we find
\begin{multline}
\label{OneLoopCorrections}
V(\phi,H)\, \supset \, \frac{\Lambda_{\mathcal{G}_F}^2 m_H^2(\phi) }{16 \pi^2} \, + \, \frac{m_H^4(\phi)}{16 \pi^2} \log\left(\frac{m_H^2(\phi)}{\Lambda_{\mathcal{G}_F}^2}\right) \, + \, \frac{\Lambda_f^4 \Lambda_{\mathcal{G}_f}^2}{16 \pi^2 M_f^2} \cos\left(\frac{\phi}{f}\right) \\
 + \, \left[ \frac{\Lambda_f^8}{16 \pi^2 M_f^4} \cos^2\left(\frac{\phi}{f}\right) \, + \, \frac{\Lambda_f^4 m_H^2(\phi) }{8 \pi^2 M_f^2} \, \cos\left(\frac{\phi}{f}\right)\right] \, \log \left(\frac{m_H^2(\phi)}{\Lambda_{\mathcal{G}_f}^2}\right) \, ,
\end{multline}
where we have neglected some subdominant terms. In the opposite region $m_H^2(\phi) \gg \Lambda_{\mathcal{G}_f}^2$ or $\Lambda_{\mathcal{G}_F}^2$, on the other hand, the corrections are strongly suppressed.\footnote{See the one-loop effective potential e.g.~in Eq.~(2.64) of Ref.~\cite{Sher:1988mj} in the limit $U''\gg \Lambda^2$.} This ensures that the term proportional to $m_H^2(\phi) \cos(\phi/f)$ gives only a small contribution to the Higgs-independent barrier. In order to guarantee that the other term proportional to $\cos(\phi/f)$ is suppressed too, we require that
\be
\label{EFTLoopCorrectionCondition1}
\Lambda_{\mathcal{G}_f} \, \lesssim 4 \pi M_f \, .
\ee
Provided that $\Lambda, \Lambda_{\mathcal{G}_F},\Lambda_F \lesssim 4 \pi M_F$ the first two terms in \Eq{OneLoopCorrections} give small corrections to the sliding term for the relaxion and do not affect the dynamics. Finally if ${\Lambda_f^2 \lesssim 4 \pi M_f v_\ew}$, the $\cos^2(\phi/f)$-term is negligible compared to the Higgs-dependent barrier when the Higgs reaches the electroweak scale. Using \Eq{EWscalerelation}, this translates to the constraint
\be
\label{EFTLoopCorrectionCondition2}
\Lambda_F \, \lesssim \, \sqrt{4 \pi} \, v_\ew  \left(\frac{F}{f}\right)^{1/4} \, .
\ee
This is less stringent than \Eq{eq:barrierValue}.

\subsection{Conditions on the warped model}
\label{sec:ConditionsFromEWBarrier}
The Higgs is localized on or near the IR brane in our warped model. Its mass parameter is then naturally of order $\Lambda_\IR^2$. We therefore identify the cutoff of our relaxion model with the IR scale:
\be
\Lambda \, \sim \, \Lambda_\IR \, .
\ee
As we have discussed in \Sec{sec:HierarchicalDecayConstants}, we can obtain the decay constants ${f_\B \approx \Lambda_\IR}$ from a Chern-Simons term in the bulk and $f_\UV \approx M_\PL^2/\Lambda_\IR$ from an effective anomalous coupling on the UV brane. Since  $F\gg f$ is required, we identify $F= M_\PL^2/\Lambda_\IR$ and $f=\Lambda_\IR$.

From the conditions in Eqs.~\eqref{ExpansionConstraint} and \eqref{eq:barrierValue} and using that $M_F \approx \Lambda_\IR$, we obtain an upper bound on the IR scale in our warped model:
\be
\label{UpperLimitLambdaIR1}
\Lambda_\IR \, \lesssim \, \left( v_\ew^2 M_\PL \right)^{1/3} \, \approx \, 4 \cdot 10^4 \,\, \text{TeV} \, .
\ee
Note that this is slightly lower than the maximal cutoff found in Ref.~\cite{Graham:2015cka}.
The reason is that there the bound on the cutoff is partly determined by the requirement of a finite viable window for the Hubble rate.
In our warped model, the corresponding contraint in \Eq{InflationConstraint} is always trivially satisfied as we discuss below. The dominant bound on the cutoff instead involves the constraint in \Eq{ExpansionConstraint} that the $H^2 \cos(\phi/F)$-term in the potential can compensate for a Higgs mass near the cutoff. This difference arises because $g$ is a free parameter in the effective description of Ref.~\cite{Graham:2015cka}, whereas in our warped model $g \propto 1/F$ is determined in terms of other parameters.

We  need to ensure that collider and flavour bounds on the KK modes in our warped model are fulfilled. We have assumed that the Standard Model fields live in the bulk. The dominant constraints then arise from $CP$-violation in $K-\bar{K}$-mixing and the electric dipole moment of the neutron.
This requires \cite{Agashe:2004cp,Csaki:2008zd}:
\be
\label{LowerLimitLambdaIR}
\Lambda_\IR \, \gtrsim \, 10 \, \text{TeV} \, .
\ee
This also satisfies constraints from electroweak precision tests without imposing a custodial symmetry \cite{Casagrande:2008hr,Bauer:2009cf} and on the radion (for a typical stabilization mechanism).

The potential leads to mixing between the Higgs and the relaxion. This further constrains the IR scale. We use results from \Ref{Flacke:2016szy}, where bounds on the parameter $\Lambda_{\rm br}^2 = \Lambda_f^2 v_\ew /M_f$ controlling the mixing have been derived from several experiments (fifth force, astrophysical and cosmological probes, beam dump, flavor, and collider searches). Using \Eq{EWscalerelation}, this translates to limits on $\Lambda_F$ and thereby on $\Lambda_\IR$. For our case $F= M_\PL^2/\Lambda_\IR$ and $f=\Lambda_\IR$, the most stringent bound comes from the distortion of the diffuse extra-galactic background light spectrum due to relaxion late decays.
This gives the constraint
\be
\Lambda_\IR \lesssim 4 \cdot 10^3 \, \text{TeV}
\ee
which is more stringent than Eq.~\eqref{UpperLimitLambdaIR1}.

\begin{table}
\begin{center}
\renewcommand{\arraystretch}{1.88}
\begin{tabular}{| c | c | c | c | c | c | c | c |}
\hline
$\Lambda$ & $F$ & $\Lambda_F$ & $M_F$ & $f$ & $\Lambda_f$ & $M_f$ &
\tabularnewline
\hline
\hline
$\Lambda_\IR$ & $\tfrac{M_\PL^2}{\Lambda_\IR}$ & $\Lambda_\IR$ & $\Lambda_\IR$ & $\Lambda_\IR$ & $\tfrac{\Lambda_\IR^{3/2}}{M_\PL^{1/2}}$ & $v_\ew$ & $10 \, \text{TeV} \lesssim \Lambda_\IR \lesssim 4 \cdot 10^3 \, \text{TeV} $
\tabularnewline [0.65ex]
\hline
\end{tabular}
\caption{\label{t:parameters}\emph{Parameters in the potential in Eq.~\eqref{eq:Vrelaxion2b} in our warped model with an electroweak-scale barrier. The range for the IR scale is allowed by all phenomenological constraints considered in this section.}}
\end{center}
\end{table}

We have discussed the confinement of $\mathcal{G}_f$ and $\mathcal{G}_F$ in terms of only the massless modes of the gauge fields in our extra-dimensional model. This is a good approximation provided that the confinement scales are smaller than the KK mass scale:\footnote{It may be possible to alleviate this condition by including some of the KK modes in the effective theory. }
\be
\label{ConfinementScaleConstraint}
\Lambda_{\mathcal{G}_f}, \Lambda_{\mathcal{G}_F}\, \lesssim \Lambda_\IR \, .
\ee
Since $\Lambda_F \lesssim \Lambda_{\mathcal{G}_F}$ and $M_F \sim \Lambda_\IR$ according to \Eq{e:dictionary}, it then follows from \Eq{ExpansionConstraint} that $\Lambda_F \sim \Lambda_\IR$ is required for successful relaxation. This in turn means that ${m_\chi,\Lambda_{\mathcal{G}_F} \sim \Lambda_\IR}$. Since the fermions $\chi,\chi^c$ are localized on the IR brane, the former condition can be naturally fulfilled. In order to discuss the latter condition, let us focus on $\mathcal{G}_F =SU(N)$ for definiteness. If we estimate the confinement scale as the scale where the 4D gauge coupling diverges, we find (see e.g.~Ref.~\cite{Csaki:2007ns})\footnote{Brane-localized kinetic terms for the gauge field would give another factor multiplying one side of this relation. This would change the required relation between $g_5^c$ and $N$ accordingly. }
\be
\label{ConfinementScale}
\frac{\Lambda_{\mathcal{G}_F}}{M_\PL} \, \approx \, \left(\frac{\Lambda_\IR}{M_\PL}\right)^{\tfrac{24 \pi^2}{11  N  (g_5^c)^2 k}} \, ,
\ee
where $g_5^c$ is the 5D gauge coupling of $\mathcal{G}_F$. From this we see that the confinement scale of $\mathcal{G}_F$ is close to the IR scale if $24 \pi^2 / (11  N  (g_5^c)^2 k) \approx 1$. This can be achieved for a wide range of values for $g_5^c$ and $N$ but clearly requires a coincidence between two parameters which are a priori not related. It may be possible to instead trigger the confinement of $\mathcal{G}_F$ by adding states on the IR brane and thereby achieve $\Lambda_{\mathcal{G}_F}\sim \Lambda_\IR$ without such a coincidence. We leave a detailed study of this question to future work.

\begin{figure}
\center
\includegraphics[scale=0.2]{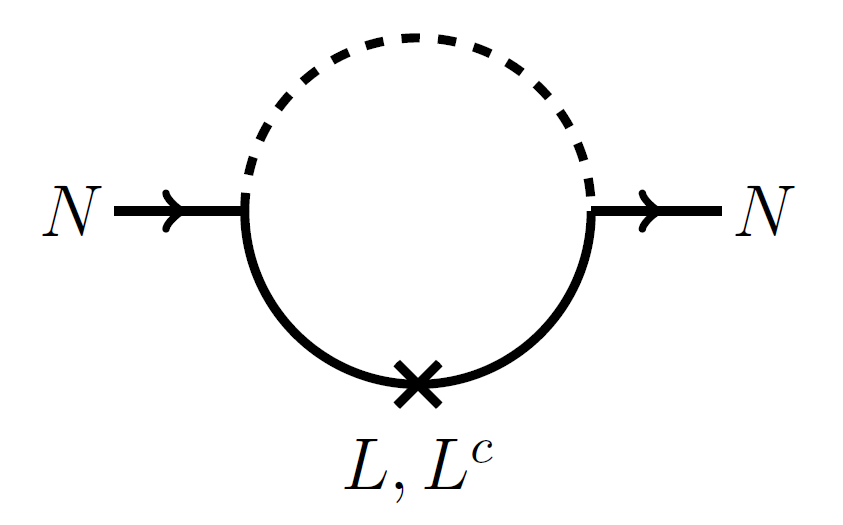}
\caption{\label{fig:loop}\emph{Loop correction to $m_N$. }}
\end{figure}

We next consider constraints related to the fermions $N,N^c$ and $L,L^c$ on the IR brane. The last two terms in Eq.~\eqref{e:mirt} break the chiral symmetry of $N,N^c$, in addition to their Dirac mass. Loop corrections then contribute to the Dirac mass (see \Fig{fig:loop}), leading to the constraint
\be
\label{mNLowerBound}
m_N \, \gtrsim \, \dfrac{y \tilde{y} \, m_L}{16 \pi^2} \,\log (\Lambda_\IR/m_L) \, .
\ee
The Higgs-dependent barrier can only stop the relaxion if $M_f \lesssim v_\ew$. Using \Eq{e:dictionary2}, the loop contribution to $m_N$ then implies that
\be
\label{mLUpperBound}
m_L \, \lesssim \,  \dfrac{4 \pi  \, v_\ew}{\sqrt{\log(\Lambda_\IR/m_L)}} \, .
\ee
The electroweak doublets $L,L^c$ can thus not be much heavier than the electroweak scale. On the other hand, due to collider constraints on such particles, they cannot be much lighter either. This limits their mass to a region near the electroweak scale. The question why their mass should be near the scale that is dynamically generated via the relaxation mechanism is the coincidence problem that we have mentioned in \Sec{sec:RelaxionPotentialFromA5}. This problem does not appear in the double-scanner scenario that we discuss in \Sec{sec:doublescanner}.

Let us briefly pause to count parameters. The potential in \Eq{eq:Vrelaxion2b} has 7 dimensionful parameters. Of these, $\Lambda$, $M_F$ and $\Lambda_F$ are of order $\Lambda_\IR$, whereas $M_f$ is of order $v_\ew$. Furthermore, $F$ and $f$ are given in terms of $\Lambda_\IR$ and $M_\PL$, while $\Lambda_f$ is fixed as a function of the other parameters via \Eq{EWscalerelation}. We can then express all parameters (up to $\mathcal{O}(1)$ factors) uniquely in terms of $\Lambda_\IR$ (plus $M_\PL$ and $v_\ew$).
In \Tab{t:parameters}, we summarize the corresponding relations and the phenomenologically viable range for the IR scale in our warped model.

Additional loop corrections arise in the effective field theory at energies below $\Lambda_{\mathcal{G}_F}$ and $\Lambda_{\mathcal{G}_f}$ as discussed in \Sec{sec:GeneralConditions}. In particular, \Eq{EFTLoopCorrectionCondition1} is an upper bound on the confinement scale of $\mathcal{G}_f$. An additional constraint arises from the requirement that the mass of the lightest fermion after diagonalizing \Eq{e:mirt} is smaller than the confinement scale (cf.~the comment above \Eq{e:potf}). Together this gives
\be
\label{eq:Lgf1}
 \abs{ m_N \, - \, \frac{y \tilde{y} \, v_\ew^2}{2 m_L} } \, \lesssim \, \Lambda_{\mathcal{G}_f} \, \lesssim \, 4 \pi \, v_\ew \, ,
\ee
where we have used $M_f \approx v_\ew$ and that the largest Higgs VEV of interest is the electroweak scale (as the relaxion stops before the Higgs VEV can grow even further). Using \Eq{EWscalerelation} and that $\Lambda_f \lesssim \Lambda_{\mathcal{G}_f}$, this upper bound on $\Lambda_{\mathcal{G}_f}$ gives an upper bound on $\Lambda_F$ which is less stringent than Eq.~\eqref{eq:barrierValue}. On the other hand, $\Lambda_{\mathcal{G}_f}$ can be very low provided that $y, \tilde{y}$ and $m_N$ are sufficiently small.  In order to ensure that $\mathcal{G}_f$   does not contribute to dark radiation during big bang nucleosynthesis, its confinement scale should be larger than a few MeV:
\be
\label{BBNconstraint}
\Lambda_{\mathcal{G}_f} \, \,\gtrsim \, \mathcal{O}(\text{few}) \cdot \text{MeV} \, .
\ee
From \Eq{EWscalerelation} and since $\Lambda_f \lesssim \Lambda_{\mathcal{G}_f}$, it follows that such low $\Lambda_{\mathcal{G}_f}$ is only possible for the IR scale near its lower bound in \Eq{LowerLimitLambdaIR}.
Furthermore, we need to ensure that the decay of composite states does not destroy heavy elements during big bang nucleosynthesis.
The resulting limits have been worked out in \Ref{Beauchesne:2017ukw}. For $\Lambda_{\mathcal{G}_f}=10 \, \text{MeV}$, ${m_L=500\, \text{GeV}}$ and $y=2\tilde{y}$, it is found that $ y,\tilde{y} \gtrsim 0.15$ is required.
This limit quickly becomes weaker for larger $\Lambda_{\mathcal{G}_f}$ or smaller $m_L$. On the other hand, the Yukawa couplings must not be too large in order to satisfy bounds on the invisible decay width of the Higgs. The corresponding limit is $ y,\tilde{y} \lesssim 0.1$ for $y=\tilde{y}$ and $m_L = 200 \, \text{GeV}$ which becomes slightly less stringent for larger $m_L$.

Given that the fermions $\chi$, $\chi^c$, $L$, $L^c$, $N$ and $N^c$ are all localized on the IR brane, we expect higher-dimensional terms in the action. These include
\be
\label{HigherDimensionalOperators}
S_{\rm 4D} \, \supset \, \int d^4x \left( c_{\chi \chi} \, \frac{m_\chi^2}{\Lambda^4_\IR} \, (\chi \chi^c)^2   \, + \, c_{N N}\frac{m_N^2}{\Lambda^4_\IR} \, (N N^c)^2
\, + \, c_{\chi N}\frac{m_\chi m_N}{\Lambda^4_\IR} \, \chi \chi^c \, N N^c \, + \, \text{h.c.} \right) \, .
\ee
The coefficients $c_{\chi \chi}$, $c_{N N}$ and $c_{\chi N}$ could be estimated using naive dimensional analysis. For simplicity, we assume them to be real. After confinement, this gives the additional terms
\be
\label{AdditionalTermsPotential}
V(\phi,H) \, \supset \, c_{\chi \chi}\frac{\Lambda_F^8}{\Lambda_\IR^4} \cos\left(\frac{2\phi}{F}\right) \,  +  \, c_{N N}\frac{\Lambda_f^8}{\Lambda_\IR^4} \cos\left(\frac{2 \phi}{f}\right) \, + \, c_{\chi N}\frac{\Lambda_F^4 \Lambda_f^4}{\Lambda_\IR^4}  \cos\left(\frac{\phi}{F}+\frac{\phi}{f}\right)
\ee
in the Higgs-relaxion potential. Note that higher-dimensional couplings involving $L L^c$ either do not directly contribute to the potential as the pair $L L^c$ does not condense or the contribution is very suppressed.\footnote{A higher-dimensional coupling $(\chi \chi^c)^\dagger N N^c$ would give a term proportional to $\cos(\phi/F - \phi/f)$ in the potential. }
The first term in \Eq{AdditionalTermsPotential} contributes to the sliding term for the relaxion. But for $c_{\chi \chi} \lesssim 1$ as expected from naive dimensional analysis, this is suppressed compared to the sliding term in \Eq{eq:Vrelaxion2b} and can thus be neglected. The second and third term, on the other hand, give additional contributions to the Higgs-independent barrier for the relaxion. Again these are suppressed compared to the barrier in \Eq{eq:Vrelaxion2b} and can be neglected. Adding higher-dimensional couplings to the Higgs in \Eq{HigherDimensionalOperators} gives terms which can similarly be neglected.

Finally, we check constraints related to inflation. Due to the temperature and quantum fluctuations in de-Sitter space, we need to demand that the confinement scales of $\mathcal{G}_f$ and $\mathcal{G}_F$ are larger than the Hubble rate during inflation:
\be
\label{InflationScaleConstraint}
H_I \, \lesssim \,  \Lambda_{\mathcal{G}_f},\Lambda_{\mathcal{G}_F} \, .
\ee
For both $\Lambda_{\mathcal{G}_F} \sim \Lambda_\IR$ and $\Lambda_{\mathcal{G}_f}\gtrsim \Lambda_f$ given by \Eq{EWscalerelation}, this is less stringent than \Eq{UpperLimitHI} from requiring that quantum fluctuations of the relaxion are negligible for the dynamics.
For $F=M_\PL^2/\Lambda_\IR$ and since $\Lambda \sim \Lambda_F \sim M_F \sim \Lambda_\IR$, the condition for having a finite viable window for the inflation scale in \Eq{InflationConstraint} is trivially fulfilled.
Furthermore, the upper limit on the inflation scale in \Eq{UpperLimitHI} is significantly smaller than the IR scale. We will assume that the inflationary sector, which we do not specify further, is located on the UV brane. Then $H_I \ll \Lambda_\IR$ guarantees that the effect of inflation on the geometry of the extra dimension is negligible \cite{Giudice:2002vh,Im:2017eju}. Similarly, for a typical stabilization mechanism it ensures that the extra dimension is safe from destabilization during inflation. In order to ensure that the barrier for the relaxion is not removed during reheating after inflation, we demand that the reheating temperature be below $\Lambda_{\mathcal{G}_f}$. This may require a relatively low reheating temperature. As follows from \Eq{BBNconstraint}, it can still be sufficiently high to allow for big bang nucleosynthesis though. Under certain conditions, the reheating temperature may also be higher than $\Lambda_{\mathcal{G}_f}$ \cite{Graham:2015cka} (see also \cite{Choi:2016kke}).

To summarize, after imposing all the constraints the usual parameters of the relaxion potential \eqref{eq:VRelaxion1a} in the model discussed in Sec.~\ref{ss:toymodel} can be written just in terms of $\Lambda_\IR$, $v_\ew$ and  $M_\PL$ as can be seen from Table 2 and using Eq.~\eqref{e:CouplingIdentification}.
The dimensionless couplings of the relaxion potential and the relaxion mass are now determined as
\begin{equation}
g\,=\,g'\,=\,\frac{\Lambda_\IR^2}{M_\PL^2}, \qquad m_{\phi} \sim \frac{ \Lambda_\IR^2}{M_\PL}\, .
\end{equation}
These couplings can thus be very small, provided that there is a large hierarchy between the IR scale and the Planck scale. This in turn can be naturally achieved (i.e.~without the input of very small numbers) e.g.~by means of the Goldberger-Wise mechanism to stabilize the extra dimension \cite{Goldberger:1999uk}. 

In addition to $\Lambda_\IR$ and $M_\PL$, the input parameters of the model discussed in Sec.~\ref{ss:toymodel} include the confinement scales $\Lambda_{\mathcal{G}_F}$ and $\Lambda_{\mathcal{G}_f}$, the fermion masses $m_\chi$, $m_N$ and $m_L$ and the couplings $y$ and $\tilde{y}$. Of these, $\Lambda_{\mathcal{G}_F}$ and $m_\chi$ are both required to be of order the IR scale. Since the corresponding fermions are localized on the IR brane, the former condition can be naturally fulfilled, while the latter condition may require a coincidence of parameters as discussed around Eq.~\eqref{ConfinementScale}. After imposing this, the electroweak scale is determined by $\Lambda_{\mathcal{G}_f}$, $y$, $\tilde{y}$, $m_N$ and $m_L$ (plus $\Lambda_\IR$ and $M_\PL$) as follows from Eqs.~\eqref{e:dictionary2} and \eqref{EWscalerelation}.
Using Eq.~\eqref{mNLowerBound} and the requirement that $M_f \lesssim v_\ew$ as well as imposing that $m_L \gtrsim v_\ew$ to satisfy electroweak precision tests \cite{Beauchesne:2017ukw}, we see that
\begin{gather}
\label{mLconditions}
v_\ew \, \lesssim \,  m_L \, \lesssim \,  \frac{4 \pi \, v_\ew}{\sqrt{\log(\Lambda_\IR /v_\ew)}} \\
\label{mNconditions}
y \tilde{y} \, m_L \, \frac{\log(\Lambda_\IR/m_L)}{16 \pi^2} \, \lesssim \,  m_N \, \lesssim \, y \tilde{y} \, \frac{v_\ew^2}{m_L}  \, . 
\end{gather}
Using the range for $m_L$ in the range for $m_N$, we then find
\be 
\label{mNRange}
y \tilde{y} \, v_\ew \, \frac{\log(\Lambda_\IR/v_\ew)}{16 \pi^2} \, \lesssim \,  m_N \, \lesssim \, y \tilde{y} \, v_\ew \, .
\ee
The fact that the electroweak doublets need to be close to the electroweak scale is the coincidence problem discussed after Eq.~\eqref{mLUpperBound}. Note that the condition for the mass of the singlets can be naturally fulfilled if it dominantly arises from the loop process in Fig.~\ref{fig:loop} (cf.~Eq.~\eqref{mNLowerBound}).
Demanding that the right electroweak scale is obtained, we then see from Eq.~\eqref{EWscalerelation} that
\be 
\label{LambdaGfRelation}
 \Lambda_{\mathcal{G}_f}^3 \, \approx \, \frac{m_L}{y \tilde{y} \, v_\ew^2} \frac{\Lambda_\IR^6}{M_\PL^2}\, ,
\ee
where $y$ and $\tilde{y}$ need to be chosen such that Eqs.~\eqref{eq:Lgf1} and \eqref{BBNconstraint} for $\Lambda_{\mathcal{G}_f}$ as well as the limits discussed below Eq.~\eqref{BBNconstraint} are fulfilled. 

In the inflationary sector, the allowed window of Hubble scales and the minimum number of e-folds are given by
\begin{equation}
\frac{\Lambda_\IR^{2}}{M_\PL} \lesssim H_I \lesssim  \frac{\Lambda_\IR^{5/3}}{M_\PL^{2/3}}\,, \qquad \mathcal{N}_e \, \gtrsim \frac{M_\PL^2}{\Lambda_\IR^{2}}\,.
\end{equation}

In Table~\ref{t:numparameters}, we give numerical values 
for two benchmark points. For the first one, we set the cutoff to its maximal allowed value in our model, $\Lambda_\IR=4 \cdot 10^3\,$TeV, and choose $y=2\tilde{y}=0.2$ and $m_L=700\,$GeV. For the second one, we choose the intermediate cutoff $\Lambda_\IR=500\,$TeV as well as $y=2 \tilde{y}=0.04$ and $m_L=450 \,$GeV. For both benchmark points, we assume that $m_N$ is dominantly generated by the loop process in Fig.~\ref{fig:loop} in which case the lower bound in  Eq.~\eqref{mNconditions} is saturated (while our choices for $m_L$ satisfy the bound in Eq.~\eqref{mLconditions}). This in particular leads to $M_f\sim v_\ew$ as used for Table \ref{t:parameters}. Both benchmark points satisfy the constraints in Eqs.~\eqref{eq:Lgf1} and \eqref{BBNconstraint} in addition to the relevant constraints from colliders and big bang nucleosynthesis as can be seen from Fig.~10 in Ref.~\cite{Beauchesne:2017ukw}. Note that for cutoffs $\Lambda_\IR \lesssim 500\,$TeV, constraints from big bang nucleosynthesis can become problematic. Indeed from Eqs.~\eqref{mNRange} and \eqref{LambdaGfRelation} and the requirement that $m_N \lesssim \Lambda_{\mathcal{G}_f}$, we 
see that lower cutoffs necessitate smaller values for $y\tilde{y}$. If $y \sim \tilde{y}$, this leads to longer lifetimes for the lightest $N N^c$ bound states which arise from the confinement of $\mathcal{G}_f$ (see \cite{Beauchesne:2017ukw}). For too long lifetimes, these decay during big bang nucleosynthesis. One way out is to choose $y\sim 1 \gg \tilde{y}$. The large coupling $y$ then allows for relatively fast decays via an off-shell $Z$ \cite{Beauchesne:2017ukw}. For example for $\Lambda_\IR=10\,$TeV, $y=1,\tilde{y}= 10^{-9}, m_L=800\,$GeV and assuming that the mass of the lightest $N N^c$ bound state is $\sim 3 \Lambda_{\mathcal{G}_f}$, we find that its lifetime is of order $1000\,$s while it can kinematically only decay into electron pairs or lighter states. This then satisfies the corresponding limit on the lifetime of order $10^4\,$s \cite{Kawasaki:2017bqm}. Alternatively, one could add new decay channels for the bound states which can allow them to decay faster and sufficiently long before big bang nucleosynthesis. We leave a further investigation of this possibility for future work.

\begin{table}
\begin{center}
\renewcommand{\arraystretch}{1.5}
\begin{tabular}{| c | c | c | c | c | c | c | c | c | c |}
\hline
$g,\,g'$ & $\frac{m_{\phi}}{\text{GeV}}$ & $\frac{\Lambda_{\mathcal{G}_f}}{\text{GeV}}$ &  $\frac{m_{N}}{\text{GeV}}$& $\frac{H_I}{\text{GeV}}$  & $\mathcal{N}_e^{\text{min}}$
\tabularnewline
\hline
\hline
$3 \cdot 10^{-24}$ & $7 \cdot 10^{-6}$ &$7 $  & $0.8$ & $[7 \cdot 10^{-6}\,,\,0.06]$   & $4 \cdot 10^{23}$\\
$4 \cdot 10^{-26}$ & $ 10^{-7}$& $0.3$  &  $2\cdot 10^{-2}$ & $[10^{-7}\,, 2 \cdot 10^{-3}]$   &  $2 \cdot 10^{25}$
\tabularnewline [0.65ex]
\hline
\end{tabular}
\caption{\label{t:numparameters}\emph{Numerical values of the parameters for two benchmark points. For the first line, we set ${\Lambda_\IR = 4\cdot 10^3\,\text{TeV}}$, $y=2\tilde{y}=0.2$ and $m_L=700\,$GeV, while for the second line, $\Lambda_\IR = 500\,$TeV, $y=2 \tilde{y}=0.04$ and $m_L=450 \,$GeV.}}
\end{center}
\end{table}

\section{Warping the  double-scanner mechanism }
\label{sec:doublescanner}

\subsection{A UV completion}

As discussed in \Sec{GeneralSetup}, the Higgs-dependent barrier in the relaxion potential needs to dominate over the Higgs-independent one once the Higgs VEV has reached the electroweak scale. This requires that $M_f \lesssim v_\ew$ which in turn necessitates to introduce new particles coupled to the Higgs near the electroweak scale. We now discuss an interesting alternative presented in \Ref{Espinosa:2015eda}. The idea is to have another axion-like scalar $\sigma$ with couplings in the potential
\begin{equation}
\label{eq:Vsigma}
V(\phi,\sigma,H) \, \supset \,  g_\sigma  \Lambda^3  \sigma \,+ \, \Lambda_\cf^4 \, \left(1 \, - \, \tilde{g}_\sigma \frac{\sigma}{\Lambda} \, + \, \tilde{g} \frac{\phi}{\Lambda}\, + \, \frac{H^2}{M_f^2} \right) \, \cos\left(\frac{\phi}{f}\right)
\end{equation}
and arrange its evolution such that it cancels off the Higgs-independent barrier. Note that we have also included a term $\phi \cos(\phi/f)$ in the potential which will be important. The remaining terms involving the relaxion are as in Eq.~\eqref{eq:VRelaxion1a}. Similar to the relaxion, the shift-symmetry breaking couplings $g_\sigma$ and $\tilde{g}_\sigma$ of the field $\sigma$ are taken to be very small.

Let us assume that $\sigma$ begins its evolution at some initial value $\sigma \gtrsim (\Lambda + \tilde{g} \phi) / \tilde{g}_\sigma$ so that the Higgs-independent term in brackets in \Eq{eq:Vsigma} is unsuppressed. Provided that ${g \Lambda^3 \lesssim \Lambda_\cf^4/f}$, the barrier term for the relaxion then dominates over its sliding term and the relaxion is initially stuck in a local minimum. Meanwhile, the first term in Eq.~\eqref{eq:Vsigma} causes $\sigma$ to slide and it eventually reaches the value $\sigma \simeq (\Lambda + \tilde{g} \phi) / \tilde{g}_\sigma$. This removes the barrier for the relaxion which can subsequently also slide down the potential. Both $\sigma$ and $\phi$ then roll down if they track each other according to the relation $\sigma \simeq (\Lambda + \tilde{g} \phi) / \tilde{g}_\sigma$.
The resulting growth of $\phi$ after a while causes the Higgs mass parameter to turn tachyonic and $H$ begins to grow too. Shortly afterwards, the Higgs-dependent barrier in Eq.~\eqref{eq:Vsigma} then becomes so big that the relaxion stops again. Provided that $\sigma$ can no longer cancel this barrier, the relaxion remains stuck. This mechanism works for certain ranges of parameters which we review below. It then allows the backreaction from the Higgs to stop the relaxion once its VEV has reached the electroweak scale even if $M_f \gg v_\ew$.

We first present a construction to generate the required terms in the potential (see also \cite{Evans:2016htp,Evans:2017bjs}). This construction is, in fact, largely independent of the embedding into warped space and can thus be used in other UV completions of the relaxion as well. It is meant to serve as a proof of principle, and does not preclude the existence of simpler or more complete models. Let us introduce an additional $U(1)$ gauge symmetry in the bulk. We identify the field $\sigma$ with the 5th component of the gauge field after imposing appropriate boundary conditions. In order to generate the sliding term in Eq.~\eqref{eq:Vsigma}, we add an anomalous coupling of $\sigma$ to a non-abelian gauge group $\mathcal{G}_{F_\sigma}$ on the UV brane using the construction in \Sec{ss:uvfermions}. We also introduce two chiral fermions $\rho$ and $\rho^c$ on the UV brane, with a Dirac mass $m_\rho$ and in respectively the fundamental and anti-fundamental representation of $\mathcal{G}_{F_\sigma}$. These fermions have no explicit coupling to $\sigma$. Such a coupling is then generated if we perform a chiral rotation of $\rho$ or $\rho^c$ to remove the anomalous coupling of $\sigma$ to $\mathcal{G}_{F_\sigma}$. If the gauge group confines at some scale $\Lambda_{\mathcal{G}_{F_\sigma}} > m_\rho$, this gives rise to the potential
\be
\label{SigmaSlidingTerm}
V(\phi,\sigma,H) \, \supset \, 2|m_\rho| \, \Lambda_{\mathcal{G}_{F_\sigma}}^3  \cos \left(\frac{\sigma}{F_\sigma} + b_\rho\right) \, .
\ee
Here $F_\sigma \gg f$ is the decay constant resulting from the anomalous coupling and $b_\rho = \arg(m_\rho)$ is the phase of the mass term. As we see later, we again have $\Lambda=\Lambda_\IR$. Expanding in $\sigma$ around the linear part of the trigonometric potential gives the sliding term in Eq.~\eqref{eq:Vsigma} with
\be
\label{gsigma}
g_\sigma \, = \, \frac{|m_\rho| \, \Lambda_{\mathcal{G}_{F_\sigma}}^3 }{F_\sigma \, \Lambda_\IR^3}
\ee
up to factors of order one.

Generating the coupling of $\sigma$ to the periodic potential for $\phi$ is somewhat more involved. Notice that in Eq.~\eqref{eq:Vsigma}, the periodic potential for $\phi$ appears with the same phase in the last four terms (which for definiteness we have chosen as $\cos (\phi/f)$). Having the same phase to a high precision in these a priori independent terms is in fact necessary for the double-scanner mechanism to work. Let us assume that $\sigma$ instead couples to $ \sin (\phi/f)$. Keeping the phases for the other periodic terms fixed, the barrier in Eq.~\eqref{eq:Vsigma} then reads
\be
V(\phi,\sigma,H) \, \supset \,   \Lambda_\cf^4 \, \left(1 \, - \, \tilde{g}_\sigma \frac{\sigma}{\Lambda} \,  \tan\left(\frac{\phi}{f}\right) \, + \, \tilde{g} \frac{\phi}{\Lambda} \, + \, \frac{H^2}{M_f^2} \right) \, \cos\left(\frac{\phi}{f}\right) \, .
\ee
Even if $\sigma$ can then initially cancel off the Higgs-independent terms (which depending on the initial value for $\phi$ may require $\sigma \gg \Lambda/\tilde{g}_\sigma$), this cancellation is generically irreversibly spoiled once $\phi$ starts rolling. The same holds for a phase difference less than $\pi/2$, if the other periodic terms have different phases or if the decay constants in the periodic terms differ from each other (in all cases down to values which are determined by the small couplings in the potential).

In order to ensure the required phase and period structure, we extend the gauge symmetry $\mathcal{G}_f$ in the bulk from Sec.~\ref{ss:toymodel} to the product group $\mathcal{G}_{f_1} \times \mathcal{G}_{f_2} \times  \mathcal{G}_{f_3} \times \mathcal{G}_{f_4} $. In addition, we impose discrete symmetries $\mathbb{Z}_2$ and $\mathbb{Z}'_2$ that interchange the groups as follows:
\be
\begin{aligned}
\label{SymmetryStructure}
 \mathcal{G}_{f_1} & \quad  \overset{\mathbb{Z}_2}{\longleftrightarrow}  \; \;  && \mathcal{G}_{f_2} \\
 \mathbb{Z}'_2 \Big \updownarrow \; \, &   &&  \, \Big \updownarrow \mathbb{Z}'_2 \\
 \mathcal{G}_{f_3} &  \quad \underset{\mathbb{Z}_2}{\longleftrightarrow} \; \; && \mathcal{G}_{f_4} \, \, .
\end{aligned}
\ee
This in particular imposes that the underlying groups (e.g.~$SU(N)$) are the same for $\mathcal{G}_{f_1}, \mathcal{G}_{f_2},\mathcal{G}_{f_3}$ and $\mathcal{G}_{f_4}$.
We couple the 5D gauge field $A_M$ that gives rise to $\phi$ to the gauge field strengths of these four groups via Chern-Simons terms as in \Sec{ss:bulkfermions}. We impose that in the resulting anomalous couplings, $\phi$ transforms as $\phi \leftrightarrow - \phi$ under $\mathbb{Z}_2$, while it is even under $\mathbb{Z}'_2$ (by choosing the coefficients $c_\B$ in \Eq{eq:cs} to transform accordingly).
This gives
\be
\label{AnomalousCoupling1}
S_{\rm 4D}  \supset  \int d^4x \, \frac{1}{16 \pi^2} \frac{\phi}{f} \, \epsilon^{\mu \nu \rho \sigma} \left( \mathrm{Tr} \left[ G^{f_1}_{\mu \nu} G^{f_1}_{\rho \sigma} \right]  -  \mathrm{Tr} \left[ G^{f_2}_{\mu \nu} G^{f_2}_{\rho \sigma} \right]  +  \mathrm{Tr} \left[ G^{f_3}_{\mu \nu} G^{f_3}_{\rho \sigma}\right]  -  \mathrm{Tr} \left[ G^{f_4}_{\mu \nu} G^{f_4}_{\rho \sigma}
\right]  \right) \, ,
\ee
where the decay constant $f \sim \Lambda_\IR$ is equal for all gauge groups by virtue of the symmetries. We also add anomalous couplings of $\sigma$ to $\mathcal{G}_{f_3}$ and $\mathcal{G}_{f_4}$ on the UV brane, using the construction in \Sec{ss:uvfermions}. We choose $\sigma$ to be even under $\mathbb{Z}_2$.
This gives
\be\label{AnomalousCoupling2}
S_{\rm 4D} \, \supset \, \int d^4x  \,   \frac{1}{16 \pi^2}  \frac{ \sigma}{\tilde{F}_\sigma} \, \epsilon^{\mu \nu \rho \sigma} \left( \mathrm{Tr} \left[G^{f_3}_{\mu \nu} G^{f_3}_{\rho \sigma}\right] \, + \,  \mathrm{Tr} \left[G^{f_4}_{\mu \nu} G^{f_4}_{\rho \sigma} \right] \right) \, ,
\ee
where the decay constant $\tilde{F}_\sigma \gg f$ is equal for the two gauge groups by virtue of the $\mathbb{Z}_2$.
We do not add corresponding couplings to $\mathcal{G}_{f_1}$ and $\mathcal{G}_{f_2}$ though. This explicitly breaks the $\mathbb{Z}'_2$ on the UV brane.

On the IR brane, we next introduce four pairs of chiral fermions $\eta_1, \eta_1^c$, $\eta_2, \eta_2^c$, $\eta_3, \eta_3^c$ and $\eta_4, \eta_4^c$ in the fundamental and anti-fundamental representation of $\mathcal{G}_{f_1}$, $\mathcal{G}_{f_2}$,  $\mathcal{G}_{f_3}$ and $\mathcal{G}_{f_4}$, respectively. The fermion pairs interchange under $\mathbb{Z}_2$ consistent with \Eq{SymmetryStructure} but we choose $\mathbb{Z}'_2$ to be explicitly broken on the IR brane too. Including Dirac masses for the pairs of chiral fermions and higher-dimensional couplings to the Higgs, this gives
\begin{multline}
\label{ActionIRBrane}
S_{\rm 4D} \, \supset \, \int d^4x \,   \, \left( m_{\eta_1}  \left[\eta_1 \eta_1^c \, + \, \eta_2 \eta_2^c \right] \, \left(1 \, + \, c_{\eta_1} \frac{H^2}{\Lambda_\IR^2}\right) \right. \\ \left. + \, m_{\eta_3} \left[  \eta_3 \eta_3^c \, + \, \eta_4 \eta_4^c \right] \, \left( 1 \, + \, c_{\eta_3} \frac{H^2}{\Lambda_\IR^2}\right) \, + \, \mathrm{h.c.} \right) \, ,
\end{multline}
where the fields are already canonically normalized and $m_{\eta_1},m_{\eta_3} \lesssim \Lambda_\IR $. The coefficients $c_{\eta_1}$ and $c_{\eta_3} $ are a priori different from each other and could be of order one or be suppressed by a loop factor.
We can now perform the chiral rotations
\be
\begin{aligned}
\label{ChiralRotationsEta}
 \eta_1  \, & \rightarrow \, e^{i \frac{\phi}{f}} \, \eta_1 \qquad &  \eta_2  \, & \rightarrow \, e^{-i \frac{\phi}{f}} \, \eta_2 \\
\eta_3  \, & \rightarrow \, e^{i \frac{\phi}{f} + i \frac{\sigma}{\tilde{F}_\sigma}  } \, \eta_3 \qquad & \eta_4  \, & \rightarrow \, e^{ -i\frac{\phi}{f} + i\frac{\sigma}{\tilde{F}_\sigma}  } \, \eta_4
\end{aligned}
\ee
while leaving $\eta_1^c$, $\eta_2^c$, $\eta_3^c$ and $\eta_4^c$ invariant. This moves $\phi$ and $\sigma$ from Eqs.~\eqref{AnomalousCoupling1} and \eqref{AnomalousCoupling2} into Eq.~\eqref{ActionIRBrane}. We assume that the gauge groups confine at energies below the IR scale. By virtue of the $\mathbb{Z}_2$ which is unbroken everywhere, the confinement scales of $\mathcal{G}_{f_1}$ and $\mathcal{G}_{f_2}$ are identical, as are those of $\mathcal{G}_{f_3}$ and $\mathcal{G}_{f_4}$. The condensates then are pairwise equal, ${\langle \eta_1 \eta_1^c \rangle = \langle \eta_2 \eta_2^c \rangle = \Lambda_{\mathcal{G}_{f_1}}^3}$ and $\langle \eta_3 \eta_3^c \rangle = \langle \eta_4 \eta_4^c \rangle = \Lambda_{\mathcal{G}_{f_3}}^3$. The resulting potential at low energies reads
\begin{multline}
\label{WarpedDoubleScannerPotential}
V(\phi,\sigma, H) \, \supset \, 4 |m_{\eta_1}| \, \Lambda_{\mathcal{G}_{f_1}}^3  \cos \left(\frac{\phi}{f}\right) \left[\cos(b_{\eta_1}) \, + \, |c_{\eta_1}| \cos(d_{\eta_1})\frac{H^2}{\Lambda_\IR^2} \right] \\
 + \, 4 |m_{\eta_3}| \, \Lambda_{\mathcal{G}_{f_3}}^3 \cos \left(\frac{\phi}{f}\right) \left[ \cos\left(\frac{\sigma}{\tilde{F}_\sigma} + b_{\eta_3} \right)\,  + \,  |c_{\eta_3}| \cos\left(\frac{\sigma}{\tilde{F}_\sigma} + d_{\eta_3} \right) \frac{H^2}{\Lambda_\IR^2} \,\right] \, ,
\end{multline}
where $b_{\eta_1} = \arg(m_{\eta_1})$, $d_{\eta_1} = \arg(m_{\eta_1} c_{\eta_1})$, $b_{\eta_3} =  \arg(m_{\eta_3})$ and $d_{\eta_3} = \arg(m_{\eta_3} c_{\eta_3})$ are given by the complex phases of the parameters. We have kept track of the phases in order to show that all terms are proportional to $\cos (\phi/f)$ without relative phase shifts as required. This is guaranteed by the $\mathbb{Z}_2$ under which $\phi \rightarrow - \phi$ and the potential is invariant. However, note that we have tacitly assumed that the fermionic condensates are real. As we have discussed at the end of \Sec{ss:toymodel} and in Appendix \ref{AppendixB}, these phases are pion-like fields and thus dynamical. Doubling the spectrum in order to ensure that the potential for these pions factorizes from the remaining potential then fixes their phases to the same value for all four condensates and leads to an additional overall minus sign in \Eq{WarpedDoubleScannerPotential}.

On the other hand, the decay constants that appear in $\cos (\phi/f)$ between the first and second line of \Eq{WarpedDoubleScannerPotential} are the same due to the $\mathbb{Z}'_2$ in the bulk. However, note that this symmetry is broken on the UV brane by the couplings for $\sigma$ in \Eq{AnomalousCoupling2}. Nevertheless we expect that this does not affect the decay constants for $\phi$ in \Eq{WarpedDoubleScannerPotential} by virtue of the non-renomalization properties of anomalous couplings (see e.g.~Ref.~\cite{Eichhorn:2012uv}). Also any such effect would be strongly suppressed since $\tilde{F}_\sigma \gg f$. We leave a detailed study of this for future work.
Furthermore, we have allowed for the masses $m_{\eta_1}$ and $m_{\eta_3}$ being different which breaks the $\mathbb{Z}'_2$ also on the IR brane. This generically leads to a different running of the gauge couplings of $\mathcal{G}_{f_1}$ and $\mathcal{G}_{f_2}$ compared to those of $\mathcal{G}_{f_3}$ and $\mathcal{G}_{f_4}$ and accordingly different confinement scales $\Lambda_{\mathcal{G}_{f_1}}$ and $\Lambda_{\mathcal{G}_{f_3}}$. However, it does not affect the decay constants for $\phi$ in \Eq{WarpedDoubleScannerPotential} either as these are defined not involving the gauge couplings of the underlying gauge groups (cf.~Eqs.~\eqref{eq:cseffective} and \eqref{eq:fB}). As follows from Eqs.~\eqref{ChiralRotationChi} to \eqref{e:potF}, it is precisely the decay constants defined in this way which determine the period of the periodic potentials. These periods are thus not affected by the differing running of the gauge couplings.
Note also that the resulting difference between the confinement scales can be made arbitrarily small for example by increasing the number of colours of the gauge groups.

\begin{table}
\begin{center}
\renewcommand{\arraystretch}{1.2}
\begin{tabular}{| c || c | c | c | c | c | c | c | c | c | c |}
\hline
 & $\chi$ & $\chi^c$ & $\eta_1$ & $\eta_1^c$ & $\eta_2$ & $\eta_2^c$ & $\eta_3$ & $\eta_3^c$ & $\eta_4$ & $\eta_4^c$ \tabularnewline
\hline
\hline
$\mathcal{G}_F$ &  $ \square$ & $\bar{\square}$ & \bf{--} &  \bf{--} &  \bf{--} & \bf{--}  &  \bf{--} & \bf{--} &  \bf{--} & \bf{--} \tabularnewline
\hline
$\mathcal{G}_{f_1}$   & \bf{--} & \bf{--} &  $\square$ & $\bar{\square}$ & \bf{--} & \bf{--} & \bf{--} & \bf{--} &  \bf{--} & \bf{--} \tabularnewline
\hline
$\mathcal{G}_{f_2}$   & \bf{--} & \bf{--} & \bf{--} & \bf{--}  &  $\square$ & $\bar{\square}$ & \bf{--} & \bf{--} &  \bf{--} & \bf{--} \tabularnewline
\hline
$\mathcal{G}_{f_3}$   & \bf{--} & \bf{--} & \bf{--} &  \bf{--}  &   \bf{--} & \bf{--} & $\square$ & $\bar{\square}$ &  \bf{--} & \bf{--} \tabularnewline
\hline
$\mathcal{G}_{f_4}$   & \bf{--} & \bf{--} & \bf{--} &  \bf{--}  &   \bf{--} & \bf{--} &  \bf{--} & \bf{--} & $\square$ & $\bar{\square}$  \tabularnewline
\hline
\end{tabular}
\caption{\label{t:matter2}\emph{Matter content on the IR brane with gauge representations for the double-scanner model.}}
\end{center}
\end{table}

We can match with the potential in \Eq{eq:Vsigma} after expanding both Eqs.~\eqref{SigmaSlidingTerm} and \eqref{WarpedDoubleScannerPotential} in $\sigma$ around regions where the corresponding trigonometric potentials are linear. Both trigonometric potentials can be in the linear part simultaneously for example for $F_\sigma \sim \tilde{F}_\sigma$ and $b_\rho - b_{\eta_3} \sim \pi$. This also ensures that the right signs in the potential are obtained. In addition to \Eq{gsigma}, we can then identify
\be
\label{ParametersDoubleScanner}
\Lambda_f^4 \, = \, |m_{\eta_1}| \, \Lambda_{\mathcal{G}_{f_1}}^3\, , \qquad M_f \, = \, \frac{\Lambda_\IR}{\sqrt{|c_{\eta_1}|}}\, , \qquad \tilde{g}_\sigma \, = \, \frac{|m_{\eta_3}| \Lambda_{\mathcal{G}_{f_3}}^3}{|m_{\eta_1}| \Lambda_{\mathcal{G}_{f_1}}^3} \, \frac{\Lambda_\IR}{\tilde{F}_\sigma}
\ee
up to factors of order one.
Notice that \Eq{WarpedDoubleScannerPotential} contains a term $ \cos(\phi/f) \cos(\sigma/\tilde{F}_\sigma) H^2$ which is not included in \Eq{eq:Vsigma}. However, provided that for example ${|m_{\eta_3}| \Lambda_{\mathcal{G}_{f_3}}^3 \approx |m_{\eta_1}| \Lambda_{\mathcal{G}_{f_1}}^3}$ and $|c_{\eta_3}|$ is somewhat suppressed compared to $|c_{\eta_1}|$, this only gives a small correction to the Higgs-dependent barrier and therefore does not affect the dynamics. Note that this would not be possible if the $\mathbb{Z}'_2$ would be unbroken on the IR brane.

As in \Sec{ss:toymodel}, we next introduce fermions $\chi$ and $\chi^c$ in the fundamental and anti-fundamental representation of a non-abelian gauge symmetry $\mathcal{G}_F$ to generate the sliding term for the relaxion and its coupling to the Higgs. These fermions also allow us to generate the term $\phi \cos(\phi/f)$ in \Eq{eq:Vsigma}. To this end, we consider the higher-dimensional operator
\be
\label{ChiEtaCoupling}
S_{\rm 4D} \, \supset \, \int d^4x \,\left(  c_{\chi \eta_1} \frac{m_\chi  m_{\eta_1}}{\Lambda_\IR^4} \, \chi \chi^c \, \bigl(\eta_1 \eta_1^c \, + \, \eta_2 \eta_2^c\bigr) \, + \, \text{h.c.} \right)
\ee
which we expect to be present since the relevant fermions live on the IR brane. The fields are already canonically normalized and $m_\chi, m_{\eta_1}\lesssim \Lambda_\IR$. The coefficient $c_{\chi \eta_1}$ is again of order one or suppressed by a loop factor. Performing the chiral rotations in Eqs.~\eqref{ChiralRotationChi} and \eqref{ChiralRotationsEta}, we find below the confinement scales
\be
S_{\rm 4D} \, \supset \, \int d^4x \, 4 |c_{\chi \eta_1}| \, \frac{|m_\chi|  \Lambda_{\mathcal{G}_F}^3  |m_{\eta_1}| \Lambda_{\mathcal{G}_{f_1}}^3}{\Lambda_\IR^4} \cos\left(\frac{\phi}{F} + b_{\chi\eta_1} \right) \, \cos\left(\frac{\phi}{f} \right) \, ,
\ee
where $b_{\chi \eta_1} = \arg(c_{\chi \eta_1} m_\chi m_{\eta_1})$. Expanding the trigonometric function of $\phi/F$ around its linear part, we can identify
\be
\label{ParametersDoubleScanner2}
\tilde{g} \, = \, |c_{\chi \eta_1}| \, \frac{|m_\chi| \Lambda_{\mathcal{G}_F}^3}{\Lambda_\IR^3 F}
\ee
up to factors of order one. Note that the coupling in Eq.~\eqref{ChiEtaCoupling} with $\eta_1 \eta_1^c, \eta_2 \eta_2^c$ replaced by $\eta_3 \eta_3^c, \eta_4 \eta_4^c$
gives an additional term $\cos (\phi/F + \sigma/\tilde{F}_\sigma) \cos(\phi/f)$ in the potential. We expect that for example for $|m_{\eta_3}| \Lambda_{\mathcal{G}_{f_3}}^3 \approx |m_{\eta_1}| \Lambda_{\mathcal{G}_{f_1}}^3$ and the corresponding coefficient $c_{\chi \eta_3}$ being somewhat suppressed compared to $c_{\chi \eta_1}$, this does not significantly affect the dynamics.

A summary of the matter content on the IR brane is given in Table \ref{t:matter2}.

\subsection{Constraints}

We have now generated all terms in the potential of \Eq{eq:Vsigma} as well as the sliding term and coupling to the Higgs of the relaxion.
In order to see if the potential parameters in Eqs.~\eqref{gsigma}, \eqref{ParametersDoubleScanner} and \eqref{ParametersDoubleScanner2} (plus Eqs.~\eqref{e:CouplingIdentification} and \eqref{e:dictionary} for $g$ and $g'$) can take on values which allow the double-scanner mechanism to work, we next discuss various constraints. We again need to ensure that the conditions discussed in \Sec{sec:GeneralConditions} are fulfilled. In particular, the Higgs VEV once the relaxion stops is as before given by \Eq{EWscalerelation}. One difference between the potential parameters for the electroweak-scale barrier and the double scanner is that $M_f \sim v_\ew$ in the former and $M_f \sim \Lambda_\IR$ in the latter. But in both scenarios, by construction the Higgs-independent barrier plays no role and therefore only the combination $\Lambda_f^2/M_f$ is relevant for the dynamics of the relaxion and Higgs. Using \Eq{EWscalerelation} to fix the Higgs VEV, we can express this combination in terms of the decay constants and $\Lambda_F$. Constraints on these parameters therefore apply for both the electroweak-scale barrier and the double scanner. We can therefore conclude that the allowed range for the IR scale  is again given by Table \ref{t:parameters}. Note that $\Lambda_f$ and $M_f$ are different from those given in the table but the combination $\Lambda_f^2/M_f$ and the other parameters in the table agree for both scenarios. In particular, we again find that $\Lambda \sim \Lambda_\IR$ and that $\Lambda_F \sim \Lambda_{\mathcal{G}_F} \sim m_\chi \sim \Lambda_\IR$ is required. On the other hand, the constraint on $\Lambda_{\mathcal{G}_f}$ in \Eq{ConfinementScaleConstraint} can always be fulfilled as follows from \Eq{eq:barrierValue}. Similarly, we see using Eqs.~\eqref{EWscalerelation}, \eqref{UpperLimitHI} and \eqref{LowerLimitLambdaIR} that the constraints in Eqs.~\eqref{BBNconstraint} and \Eq{InflationScaleConstraint} are automatically fulfilled.

There are new conditions that are specific to the double-scanner mechanism:
The fields $\phi$ and $\sigma$ track each other according to the relation $\sigma \simeq (\Lambda + \tilde{g} \phi) / \tilde{g}_\sigma$ once the barrier is sufficiently small provided that \cite{Espinosa:2015eda}
\be
\label{DoubleScannerConstraint1}
g \, \tilde{g} \, \gtrsim \, g_\sigma  \tilde{g}_\sigma \, ,
\ee
where $g$ is given by Eqs.~\eqref{e:CouplingIdentification} and \eqref{e:dictionary}. On the other hand, $\sigma$ can no longer cancel the barrier that the Higgs generates once it obtains a VEV if \cite{Espinosa:2015eda}
\be
\label{DoubleScannerConstraint2}
g \left( \tilde{g}- \frac{g}{2 \lambda} \right) \, \lesssim \, g_\sigma  \tilde{g}_\sigma
\ee
with $\lambda$ being the Higgs quartic coupling. We have $F \approx F_\sigma \approx \tilde{F}_\sigma$ since these decay constants all arise from anomalous couplings on the UV brane. Comparing Eqs.~\eqref{e:CouplingIdentification} and \eqref{ParametersDoubleScanner2}, we also see that $\tilde{g} \sim |c_{\chi \eta_1}| g$.
On the other hand, the couplings $g_\sigma$ and $\tilde{g}_\sigma$ can a priori be quite different. The gauge group $\mathcal{G}_{F_\sigma}$ that gives rise to the sliding term for $\sigma$ can in principle be localized on the UV brane. Nevertheless we should still demand that its confinement scale is below the IR scale to ensure that the effective description for $\sigma$ is valid at the energy scale where the potential is generated. In addition, we need to require that $|m_\rho|\lesssim \Lambda_{\mathcal{G}_{F_\sigma}}$. In order to study one concrete example, let us assume that $ |m_{\eta_1}| \Lambda_{\mathcal{G}_{f_1}}^3 \approx |m_{\eta_3}| \Lambda_{\mathcal{G}_{f_3}}^3$ (corresponding to $\mathbb{Z}'_2$ being only weakly broken on the IR brane).
This gives $\tilde{g}_\sigma \approx g$ and $g \gtrsim g_\sigma $. The conditions in Eqs.~\eqref{DoubleScannerConstraint1} and \eqref{DoubleScannerConstraint2} then simplify to
\be
g \, |c_{\chi \eta_1}|  \, \gtrsim \, g_\sigma \, , \, \qquad g \left(|c_{\chi \eta_1}|\, - \, \frac{1}{2 \lambda} \right) \, \lesssim \, g_\sigma \, .
\ee
This can be fulfilled for a wide range of $g_\sigma$ if $|c_{\chi \eta_1}|\lesssim 1/(2 \lambda)$. This example shows that the conditions for the double-scanner mechanism to work can be easily satisfied.

Finally, let us consider loop corrections to the potential. The double-scanner mechanism cannot remove barriers from terms like $\cos^2(\phi/f)$ \cite{Espinosa:2015eda}. Therefore these must be smaller than the Higgs-dependent barrier when the Higgs reaches the electroweak scale. For loop corrections from the Higgs, this translates to the condition $\Lambda_f^2 \lesssim 4 \pi M_f v_\ew$ and in turn to \Eq{EFTLoopCorrectionCondition2} which is less stringent than the already imposed \Eq{eq:barrierValue}.
This also means that \Eq{EFTLoopCorrectionCondition1} can always be fulfilled.
Furthermore, in addition to \Eq{ChiEtaCoupling} we expect higher-dimensional operators like
\be
\label{HigherDimensionalOperators2}
S_{\rm 4D}  \supset  \int d^4x \left( c_{\chi \chi}  \frac{m_\chi^2}{\Lambda^4_\IR} (\chi \chi^c)^2    + c_{\eta_1 \eta_1}\frac{m_{\eta_1}^2}{\Lambda^4_\IR} \left[ \left(\eta_1 \eta_1^c\right)^2 + \left(\eta_2 \eta_2^c\right)^2 \right]  + c_{\eta_1 \eta_2}\frac{m_{\eta_1}^2}{\Lambda^4_\IR} \eta_1 \eta_1^c \, \eta_2 \eta_2^c + \text{h.c.} \right)
\ee
and similar terms involving $\eta_3, \eta_3^c, \eta_4, \eta_4^c$ since the relevant fermions are all localized on the IR brane. The coefficients are again of order one or suppressed by a loop factor and are partly determined by the $\mathbb{Z}_2$. Assuming all parameters to be real for simplicity, below the confinement scales this gives
\be
\label{AdditionalTermsPotential2}
V(\phi,H) \, \supset \, 2\,c_{\chi \chi}\frac{\Lambda_F^8}{\Lambda_\IR^4} \cos\left(\frac{2\phi}{F}\right) \,  +  \, 4\, c_{\eta_1 \eta_1} \frac{\Lambda_f^8}{\Lambda_\IR^4} \cos\left(\frac{2 \phi}{f}\right)  \, .
\ee
The first term gives a correction to the sliding term for the relaxion which is negligible for $c_{\chi \chi} \lesssim 1$. The second term, on the other hand, gives another type of barrier that cannot be cancelled by the double-scanner mechanism. It is sufficiently suppressed compared to the Higgs-dependent barrier provided that $\Lambda_f^2 \lesssim v_\ew \Lambda_\IR^2 / (M_f \sqrt{c_{\eta_1 \eta_1}})$. This in turn leads to a condition which for example for $c_{\eta_1 \eta_1} \sim c_{\eta_1} \sim 1$ is the same as \Eq{eq:barrierValue} and which is then fulfilled for the entire range of IR scales in Table \ref{t:parameters}.

\section{Conclusions}
\label{sec:conclusion}

We have implemented the cosmological relaxation mechanism in a warped extra dimension. The relaxion potential trades the hierarchy between the Planck and electroweak scale for a technically natural hierarchy of decay constants. Warped extra dimensions are then a natural choice for its UV completion as they can generate a large hierarchy of scales purely from geometry. In our construction, the relaxion is identified with the scalar component of an abelian gauge field in the bulk, whose profile automatically has a small overlap with the UV brane. The warping generates the hierarchy from the Planck scale down to the scale of the IR brane, which is then identified with the cutoff $\Lambda$ of the relaxion potential. From there onwards, the Higgs mass is relaxed down to its physical value.

In \Sec{sec:HierarchicalDecayConstants}, we have presented a model-building toolkit for generating anomalous couplings of the relaxion to new, strong sectors. Depending on the localization of the anomalous terms in the warped interval, hierarchically different decay constants for these couplings may be obtained, including decay constants which are super-Planckian.
A benchmark model coupling the relaxion to the Higgs was constructed in \Sec{sec:RelaxionPotentialFromA5}. The sliding term and its coupling to the Higgs is generated through the condensation of a Dirac pair of SM singlet fermions that live on the IR brane. The barrier term, on the other hand, is generated close to the electroweak scale by the condensation of vector-like fermions with the same quantum numbers as one generation of SM leptons. These are also localized at the IR brane, and have masses near or below the weak scale, but are a priori unrelated to it, leading to the well-known coincidence problem. In order to avoid this and achieve a larger scale for the barrier term, a more elaborate construction is required.
In \Sec{sec:doublescanner}, we have presented a warped UV completion for one such scenario, the double-scanner mechanism of \Ref{Espinosa:2015eda}.

The constraints for the model, both in general and those specific to the construction of \Sec{sec:RelaxionPotentialFromA5}, were discussed thoroughly in \Sec{sec:constraints}, as well as the stability of the potential under radiative corrections. The requirement of obtaining the correct Higgs VEV may be used to fix the scale where the barrier term is generated in terms of the other parameters. Then, we have found that the scale where the sliding and scanning terms are generated needs to be of order the IR scale. Since the SM fields live in the bulk, standard flavor constraints of Randall-Sundrum models push the minimum value of the IR scale to ${\Lambda \gtrsim 10 \, \text{TeV}}$.
The maximum cutoff that we can achieve while ensuring that all theoretical and phenomenological constraints are fulfilled is $\Lambda \approx 4 \cdot 10^6$ GeV.

In this work, we have focused on inflation to provide a friction term for the slow-roll of the relaxion, but interesting alternatives such as the particle-production mechanism of \Ref{Hook:2016mqo} exist. It would be interesting to explore how such constructions may be implemented in warped space.
The framework that we have described naturally allows for hierarchical decay constants for axion-like fields to be generated. As such it presents many further opportunities for model building, not limited to relaxion models, such as applications to inflation or dark matter.  Another interesting possibility for generating this hierarchy is to consider a more general geometry with more than one AdS$_5$ throat \cite{Fonseca:2018}.

\section*{Acknowledgments}
LdL thanks DESY for hospitality during his stay, where part of this work was completed and acknowledges support by the S\~{a}o Paulo Research Foundation (FAPESP) under grants 2012/21436-9 and 2015/25393-0. BvH thanks Fermilab for hospitality while part of this work was completed. This visit has received funding/support from the European Union’s Horizon 2020 research and innovation programme under the Marie Sk\l{}odowska-Curie grant agreement No 690575. BvH also thanks the Fine Theoretical Physics Institute at the University of Minnesota for hospitality and partial support. The work of CSM was supported by the Alexander von Humboldt Foundation, in the framework of the Sofja Kovalevskaja Award 2016, endowed by the German Federal Ministry of Education and Research. The authors would like to thank Aqeel Ahmed, Enrico Bertuzzo, Zackaria Chacko, Giovanni Grilli di Cortona, Adam Falkowski, Gero von Gersdorff, Tony Gherghetta, Christophe Grojean, Roni Harnik, Arthur Hebecker, Ricardo D.~Matheus,  Enrico Morgante, Eduardo Pont\'on, Pedro Schwaller, Marco Serone,  G\'eraldine Servant, Alexander Westphal and Alexei Yung for useful discussions and comments.

\appendix

\section{An anomalous coupling on the UV brane from two throats}
\label{Appendix1}

The interaction in \Eq{eq:intlocalizedUV} should be understood as an effective coupling that can for example arise from a Chern-Simons term in a second throat as we now briefly discuss. More details will be presented in \cite{Fonseca:2018}. To this end, we consider a setup with two warped spaces which are glued together at a common UV brane but each slice  still has its own IR brane.
For simplicity, we assume that both slices have the same AdS scale $k$. Let us denote the coordinates along the extra dimension in the two throats as $z_1$ and $z_2$, with metric in each throat again given by \Eq{e:metric}. The coordinates match at the common UV brane at $z_{\UV_1} =z_{\UV_2} =1/k$, while the IR branes are at $z_{\IR_1} = e^{k L_1}/k$ and $z_{\IR_2} = e^{k L_2}/k$. We then introduce an abelian gauge boson which propagates in both throats (see e.g.~\cite{Cacciapaglia:2005pa}). We break the gauge symmetry on the two IR branes by imposing the boundary conditions in \Eq{e:bcs} but leave it unbroken on the UV brane. This allows for one massless mode from $A_5$ which lives in both throats with wavefunction $A_5 = \mathcal{N} a(z_i)^{-1} \phi$ in a given throat (the wavefunction is continuous at the UV brane). We will be interested in the case where one throat is significantly longer than the other. The normalization constant $\mathcal{N}$ is then dominated by the longer throat. Choosing $L_1 > L_2$  without loss of generality, we have $z_{\IR_1} \gg z_{\IR_2}$, which gives $\mathcal{N} \simeq g_4 \sqrt{2 k L_1} e^{-k L_1}$ with $g_4$ defined as before. Let us next introduce a Chern-Simons coupling of $A_M$ to a non-abelian gauge group, where we choose the coupling to be localized in the second throat:
\begin{align} \label{eq:css}
S_{\rm{5D}} \, \supset\, \int d^4x \int_{z_{\UV}}^{z_{\IR_2}}\!\!dz_2\, \frac{c_{b_2}}{16 \pi^2} \epsilon^{MNPQR} A_M \, \mathrm{Tr}  \left[ \bold{G}_{NP} \bold{G}_{PQ} \right].
\end{align}
Notice that the coupling to $A_5$ from this resembles Eq.~\eqref{eq:intlocalizedUV} with the $\delta$-function replaced by the integral over $A_5$ in the second throat.
In the limit of a very short second throat with $z_{\IR_2} \sim \mathcal{O}(\text{few}) \cdot z_\UV$, we can think of this integral as a smeared-out $\delta$-function. Correspondingly we expect the decay constant of $\phi$ in this limit to agree with Eq.~\eqref{eq:fUV}. Let us again restrict ourselves to the zero-mode of the non-abelian gauge field. Integrating over the extra dimension, we in particular find
\begin{equation}
S_{\rm 4D} \, \supset \,  \int d^4x \frac{1}{16 \pi^2} \frac{ \phi(x)}{f_{\B_2}}~ \epsilon^{\mu \nu \rho \sigma} \mathrm{Tr} \left[ G_{\mu \nu} G_{\rho \sigma} \right]
\end{equation}
with decay constant given by
\be
f_{\B_2} \, \simeq \, \frac{k \, e^{k L_1- 2 k L_2}}{c_{b_2} \,g_4 \sqrt{2 k L_1} }
\ee
or $f_{\B_2}\approx \Lambda_{\IR_2}^2/\Lambda_{\IR_1}$. For a very short second throat with $L_1 \gg 2 L_2$, this indeed agrees with Eq.~\eqref{eq:fUV}. On the other hand, the two-throat construction allows for more general choices for the decay constant, with a continuum between $M_\PL^2/\Lambda_{\IR_1}$ and $\Lambda_{\IR_1}$ (as $\Lambda_{\IR_1} < \Lambda_{\IR_2}$ by assumption). The resulting phenomenology and the details of the construction will be presented in \cite{Fonseca:2018}.

\section{Chern-Simons terms from bulk fermions}
\label{AppendixA}

In this appendix, we briefly review how charged bulk fermions can give rise to Chern-Simons terms. We consider a bulk fermion $\Psi$ which couples to both the non-abelian gauge group and the $U(1)$ from \Sec{ss:bulkfermions}. The action reads
\be
S_{\rm 5D} \, \supset \, \int d^4x dz \, \sqrt{g} \, \left(\bar{\Psi}i \slashed{D} \Psi \, + \, m_\Psi \bar{\Psi} \Psi \right) \, ,
\ee
where the covariant derivative is $D_M = \partial_M - i \bm{G}_M - i A_M$ with $\bm{G}_M$ being the non-abelian gauge field (and $A_M$ the $U(1)$ gauge field). In order to see that this gives the same anomaly as a Chern-Simons term, we can perform a field redefinition \cite{Csaki:2009re, Bunk:2010gb}
\be
\Psi \, \rightarrow \exp \left[ i \int_{z_0}^z \hspace{-.2cm} d \tilde{z} A_5(x,\tilde{z}) \right] \Psi \, ,
\ee
where the constant $z_0$ can be chosen according to convenience.
However, the field redefinition is anomalous on the branes\footnote{We note that, e.g.~for $SU(N)$, there is an additional $SU(N)^3$ anomaly. It can be canceled by adding another bulk fermion, uncharged under $U(1)$, with opposite boundary conditions from $\Psi$.} and transforms the action into (see \cite{Hirayama:2003kk,ArkaniHamed:2001is,Scrucca:2001eb,Gripaios:2007tk})
\be
\label{AnomalousTransformation2}
S_{\rm 5D}  \rightarrow  S_{\rm 5D}  +  \int d^4x dz   \left( \int_{z_0}^z \hspace{-.2cm}d\tilde{z} A_5(x,\tilde{z})\right) \frac{\epsilon^{\mu \nu \rho \sigma}}{48 \pi^2} \mathrm{Tr} \left[ \bm{G}_{ \mu \nu} \bm{G}_{\rho \sigma} \right] \bigl(\alpha_\UV \delta(z-z_{_\UV}) + \alpha_\IR \delta(z-z_\IR)\bigr).
\ee
The coefficients $\alpha_\UV$ and $\alpha_\IR$ depend on the boundary conditions on the two branes for the left-handed component $\Psi_L$ of the bulk fermion (which in turn fixes the boundary conditions of the right-handed component $\Psi_R$). If $\Psi_L$ is even (odd) on a given brane, $\alpha=1 (-1)$. Let us first assume $\alpha_\UV = - \alpha_\IR$ in which case $\Psi$ does not have a massless mode.
From \Eq{AnomalousTransformation2}, we then get the anomalous coupling of $\phi$ in \Eq{eq:cseffective} with
\be
c_\B \, = \, \frac{\alpha_\IR}{4} \, .
\ee
Notice that this is independent of $z_0$. In the opposite case $\alpha_\UV = \alpha_\IR$, on the other hand, $c_\B$ depends on $z_0$. But then $\Psi$ has a massless mode which contributes to the anomaly and which cancels the dependence on $z_0$. If the Chern-Simons term arises from such a bulk fermion, any perturbative contribution to the potential for $A_5$ can be sufficiently suppressed by making the bulk mass of the fermion somewhat larger than the AdS scale (see e.g.~\cite{Contino:2003ve,Pilo:2003gu}).

\section{Pion-like fields in the relaxion potential}
\label{AppendixB}

In this appendix, we include the pion-like fields which arise from the condensing fermions on the IR brane and which contribute to the potential.
Let us focus on $\chi, \chi^c$ for definiteness. As usual, we can parametrize the pseudo-Nambu-Goldstone boson corresponding to the breaking of the chiral symmetry of $\chi, \chi^c$ by the $\sigma$-model field $U = \exp(i \pi_\chi / f_\chi)$ with a decay constant of order $f_\chi \sim \Lambda_{\mathcal{G}_F}$. After confinement then $\langle \chi \chi^c \rangle= \Lambda_{\mathcal{G}_F}^3 U$. From \Eq{S4ChiF}, this gives
\be
V(\phi,H) \, \supset \, m_\chi \, \Lambda_{\mathcal{G}_F}^3 \left( 1 \, + \, \frac{H^2}{\Lambda_\IR^2}\right)  \, \cos\left(\frac{\phi}{F} + \frac{\pi_\chi}{f_\chi}\right) \, ,
\ee
where for simplicity we again ignore phases and prefactors. Since $F \gg f_\chi$, generically $\pi_\chi$ settles into its minimum $\pi_\chi^{\rm min} = f_\chi \pi - f_\chi \phi/F $ first after which the potential becomes independent of $\phi$. This problem is remedied for example by introducing another pair of chiral fermions $\tilde{\chi} \tilde{\chi}^c$ with the same quantum numbers. Instead of \Eq{e:mir} we then have
\be
\label{e:mir2}
S_{\rm 4D} \, \supset \, \int d^4x \,  \left( 1 + \frac{H^2}{\Lambda_\IR^2} \right) \, \left[ m_\chi \, \chi \chi^c \, + \, m_{\tilde{\chi}} \, \tilde{\chi} \tilde{\chi}^c \right] \, + \, \text{h.c.}\, .
\ee
Similar to the up and down quark in the Standard Model, the fermions transform under an approximate $SU(2)_L \times SU(2)_R$ symmetry which is spontaneously broken to a diagonal $SU(2)_V$ by the condensates and explicitly but weakly broken by their masses. The corresponding pseudo-Nambu-Goldstone bosons are parametrized as
\be
U = e^{i \Pi_\chi / f_\chi} \qquad \text{with} \, \, \, \, \Pi_\chi \, = \,
\left(
\begin{matrix}
& \pi_\chi^0 & \sqrt{2} \pi_\chi^{+} \\
& \sqrt{2} \pi_\chi^{-} & -\pi_\chi^0
\end{matrix}
\right) \, .
\ee
We next perform the chiral rotation
\be
\chi \, \rightarrow \, e^{i \frac{\phi}{2F}} \chi\, ,  \, \quad \quad \tilde{\chi} \, \rightarrow \, e^{i \frac{\phi}{2F}} \tilde{\chi}
\ee
with $\chi^c$ and $\tilde{\chi}^c$ left invariant to remove the coupling of $\phi$ to $\mathrm{Tr} \left[G^{F}_{\mu \nu} G^{F}_{\rho \sigma}\right]$ in \Eq{e:f}. For this choice of chiral rotation, no kinetic mixing between the relaxion and the pions is induced (see \Ref{Georgi:1986df}). Choosing $m_\chi= m_{\tilde{\chi}}$ for simplicity, from \Eq{e:mir2} we get below the confinement scale
\be
V(\phi,H) \, \supset \, m_\chi \, \Lambda_{\mathcal{G}_F}^3 \left( 1 \, + \, \frac{H^2}{\Lambda_\IR^2}\right)  \, \cos\left(\frac{\phi}{2F} \right) \, \cos\left(\frac{\pi_\chi}{f_\chi} \right)\, ,
\ee
where $\pi_\chi \equiv \sqrt{(\pi_\chi^0)^2+2\pi_\chi^{+}\pi_\chi^{-}}$.
The potential for the pions and relaxion thus factorizes and no longer vanishes once the pions settle into their minimum.
This is similar to what happens for the axion and the pion of the Standard Model, see \Ref{diCortona:2015ldu}. For the generalization of the potential to the case $m_\chi\neq m_{\tilde{\chi}}$, see also \Ref{diCortona:2015ldu}. The potential after minimization with respect to the pion then still leads to a nonvanishing potential for the relaxion but the latter is no longer a simple cosine.

\bibliographystyle{JHEP}
\bibliography{WRelBib}

\end{document}